\documentclass[11pt,a4paper]{article}
\usepackage{epsfig}
\usepackage{fleqn}
\usepackage{jheppub}
\usepackage{amsmath}
\usepackage{latexsym}
\usepackage{array}
\usepackage[numbers,sort&compress]{natbib}
\pdfoutput=1
\title{ Measurement of the $e^+e^- \to\pi^+\pi^- $ process 
cross section with the SND detector at the VEPP-2000
collider in the energy region $0.525<\sqrt{s}<0.883$ GeV}
\author[a,b]{M.~N.~Achasov}
\author[a]{A.~A.~Baykov}
\author[a,c]{A.~Yu.~Barnyakov}
\author[a,b]{K.~I.~Beloborodov}
\author[a,b]{A.~V.~Berdyugin}
\author[a,b]{D.~E.~Berkaev}
\author[a]{A.~G.~Bogdanchikov}
\author[a]{A.~A.~Botov}
\author[a,b]{T.~V.~Dimova}
\author[a,b]{V.~P.~Druzhinin}
\author[a]{V.~B.~Golubev}
\author[a,b]{L.~V.~Kardapoltsev}
\author[a,b]{A.~G.~Kharlamov}
\author[a,b,c]{I.~A.~Koop}
\author[a,b]{A.~A.~Korol}
\author[a]{D.~P.~Kovrizhin}
\author[a,b]{A.~S.~Kupich}
\author[a]{ A.~P.~Lysenko}
\author[a]{K.~A.~Martin}
\author[a]{N.~A.~Melnikova}
\author[a,b]{N. Yu. Muchnoi}
\author[a]{A.~E.~Obrazovsky}
\author[a]{E.~V.~Pakhtusova}
\author[a,b]{E.~A.~Perevedentsev}
\author[a,b]{K.~V.~Pugachev}
\author[a,b]{Y.~S.~Savchenko}
\author[a,b]{S.~I.~Serednyakov}
\author[a,b]{Z.~K.~Silagadze}
\author[a,b]{P.~Yu.~Shatunov}
\author[a,b]{Yu.~M.~Shatunov}
\author[a]{D.~A.~Shtol}
\author[a,b]{D.~B.~Shwartz}
\author[a]{I.~K.~Surin}
\author[a]{ Yu.~V.~Usov}
\author[a,b]{I. M. Zemlyansky}
\author[a]{V.~N.~Zhabin}
\affiliation[a]{Budker Institute of Nuclear Physics, Siberian Branch of the Russian Academy of Sciences,\\
11, Acad. Lavrentiev Pr., Novosibirsk, 630090, Russia}
\affiliation[b]{Department of physics, Novosibirsk State University,\\ 1, Pirogova str., Novosibirsk, 630090, Russia}
\affiliation[c]{Novosibirsk State Technical University,\\
20 Prospekt K. Marksa, Novosibirsk,630073, Russia}
\emailAdd{a.s.kupich@inp.nsk.su}

\date{}
\abstract{The cross section of the process $e^+ e^-\to\pi^+\pi^-$ has been measured
in the Spherical Neutral Detector (SND) experiment at the VEPP-2000 $e^+e^-$ 
collider VEPP-2000 in the energy region $525 <\sqrt[]{s} <883$ MeV. 
The measurement is based on data with an integrated luminosity of about
4.6 pb$^{-1}$. The systematic uncertainty of the cross section determination is 
0.8\% at $\sqrt{s}>0.600$ GeV. The $\rho$ meson parameters are
obtained as
$m_\rho = 775.3\pm 0.5\pm 0.6$ MeV, $\Gamma_\rho = 145.6\pm 0.6\pm 0.8$ MeV,
$B_{\rho\to e^+ e^-}\times B_{\rho\to\pi^+\pi^-} = 
(4.89\pm 0.02\pm 0.04)\times 10^{-5}$, and the parameters of the 
$e^+ e^-\to\omega\to\pi^+\pi^-$ process, suppressed by $G$-parity, as
$B_{\omega\to e^+ e^-}\times B_{\omega\to\pi^+\pi^-}=
(1.32\pm 0.06\pm 0.02)\times 10^{-6} $ and 
$\phi_{\rho\omega} = 110.7\pm 1.5\pm1.0$ degrees.}
\arxivnumber{2004.00263}
\begin{document}
\maketitle
\flushbottom

\section{Introduction}

SND \cite{snd, snd-2} is a general purpose nonmagnetic detector operating at the
VEPP-2000 $e^+e^-$ collider in the center-of-mass energy range from 
0.2 to 2.0 GeV \cite{vepp2k}. Experimental studies include measurements of
the cross sections of the $e^+ e^-$ annihilation processes into hadrons.
These measurements are largely motivated by the need for high-precision
calculation of the hadronic contribution to the anomalous magnetic
moment of the muon $(g-2)/2$ \cite{g-2}. In particular, the
$e^+e^-\to\pi^+\pi^-$ cross section in the energy region below 1 GeV gives 
the dominant contribution to this value and should be measured with accuracy 
better than 1\% \cite{cs2p}.

The cross section of the $e^+e^-\to\pi^+\pi^-$ process in the energy region
$\sqrt{s}<1000$ MeV can be described within the vector meson dominance model
(VMD) framework and is determined by the transitions $V\to\pi^+\pi^-$ of the
light vector mesons ($V=\rho,\omega,\rho^\prime,\rho^{\prime\prime}$). The main contribution in this energy region comes from the 
$\rho\to\pi^+\pi^-$ and from the G-parity violating $\omega\to\pi^+\pi^-$ 
transitions. Studies of the $e^+e^-\to\pi^+\pi^-$ reaction allow us to 
determine the $\rho$ and $\omega$ meson parameters, provide information on 
the $G$-parity violation mechanism and $\rho,\rho^\prime,\rho^{\prime\prime}$ 
mixing \cite{kozev}.

The process $e^+ e^-\to\pi^+\pi^-$ in the energy region $\sqrt{s} $ below
1000 MeV was studied for more than 40 years in a number of experiments
 \cite{augu, augu-2, augu-3, ausl, ausl-2, bena, quen, vas1, buki, vas2, vas3, kur1, kur2, spec,
olya, kmd2, kmd2-1, kloe, snd2pi, snd2pi-2, kmd2-2, kmd2-3, kloe-2, kloe-3, kloe-4, babar, babar-2, bes3}.
This work presents the results of the $e^+ e^-\to\pi^+\pi^-$ cross section measurements with SND detector in the energy region
$ 525 <\sqrt{s} <883 $ MeV based on $ IL = 4.6 $ pb $^{-1} $ experimental 
data collected by SND in 2012--2013. Approximately $2.3\times 10^6 $ collinear
events are used in the analysis. About $10^6$ are events of the processes 
$e^+e^-\to\pi^+\pi^-$, $e^+e^-\to\mu^+\mu^-$ and $1.3\times 10^6$ are 
$e^+e^-\to e^+e^-$ events.

\section{Experiment}
 
\begin{figure} [tbp]
\centering
\includegraphics [width = 1.0\textwidth]{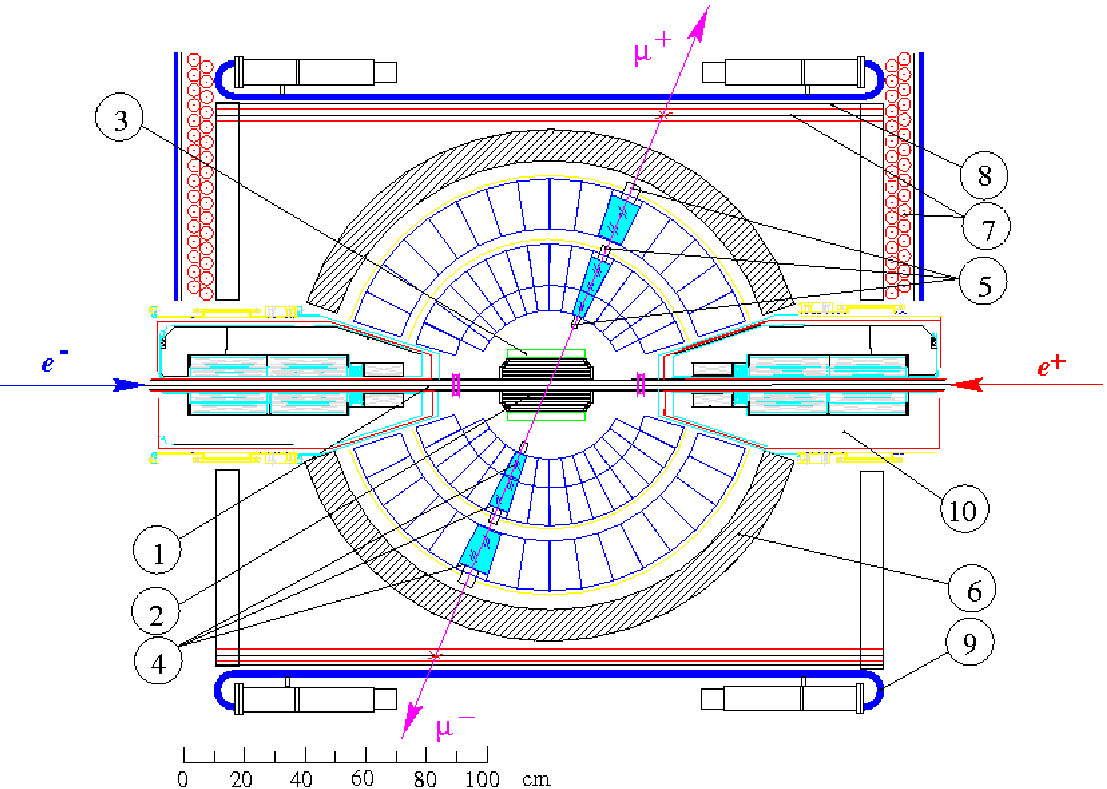}
\caption{SND detector, section along the beams: (1) beam pipe,
(2) tracking system, (3) aerogel Cherenkov counters, (4) NaI (Tl) crystals,
(5) vacuum phototriodes, (6) iron absorber, (7) proportional tubes, (8) iron
absorber, (9) scintillation counters, (10) solenoids of collider.}
\label{sndt}
\end{figure}

The SND is operated at the VEPP-2000 collider since 2010 till present day. 
It consists of a tracking system based on cylindrical drift and
proportional chambers placed in a common gas volume, aerogel threshold
counters \cite{ashif}, a three-layer spherical electromagnetic
calorimeter based on NaI (Tl) crystals and a muon system which includes
two layers of proportional tubes and scintillation counters (figure~\ref{sndt}). 
The calorimeter energy and angular resolutions depend on the photon
energy $E$ as $\sigma_E/E (\%) ={4.2 \% /\sqrt [4]{E (\mbox{GeV})}} $ and
$\sigma_{\phi,\theta} ={0.82^\circ /\sqrt []{E (\mbox{GeV})}}\oplus
0.63^\circ$. Its total solid angle is 95\% of $4\pi$.
The solid angle of the tracking system is 94 \% of $4\pi$. Its angular
resolution is $0.45^\circ$ and $0.8^\circ$ for the azimuthal and
polar angles, respectively. The threshold Cherenkov counters
are based on aerogel with the refractive index of 1.05. The threshold
momenta for $e/\mu/\pi$ are approximately equal to 1.6~/~330~/~436 MeV/c,
respectively. This system covers 60\% of the total solid angle.

The VEPP-2000 collider beam energy is determined using a 
beam-energy-measurement system based on the Compton back-scattering of laser 
photons on the electron beam. The accuracy of the beam-energy measurement is
about 30 keV \cite{ems,ems-2}.
		 
\section{Analysis}

The cross section of the process $ e^+ e^-\to\pi^+\pi^- $ is measured as 
follows.
\begin{enumerate}
\item
The collinear $e^+e^-\to e^+e^-,\pi^+\pi^-,\mu^+\mu^-$ events are selected.
\item
The selected events are sorted into the two classes: $e^+e^-$ and
$\pi^+\pi^-,\mu^+\mu^-$ using the energy depositions in the calorimeter crystals. 
\item
The luminosity is determined from the number of $e^+ e^-\to e^+ e^-$ events:
\begin{gather}
IL =\frac{N_{ee}}{\varepsilon_{ee}\sigma_{ee}}.
\end{gather}
Here $N_{ee}$, $\varepsilon_{ee}$ and $\sigma_{ee}$ are the number of 
events, detection efficiency and cross section of the process 
$e^+ e^-\to e^+e^-$ respectively. 
To obtain the number of $e^+ e^-\to\pi^+\pi^-$ events, 
the number of $e^+ e^-\to\mu^+\mu^-$ events is calculated using 
theoretical cross section as
\begin{gather}
N_{\mu\mu} = IL\varepsilon_{\mu\mu}\sigma_{\mu\mu}
\end{gather}
and then subtracted from the total number of $\pi^+\pi^-$ and $\mu^+\mu^-$ events.
Here $\varepsilon_{\mu\mu} $ and
$\sigma_{\mu\mu} $ are the detection efficiency and cross section of
$ e^+ e^-\to\mu^+\mu^- $, respectively.
\item
The Born cross section of the process $ e^+ e^-\to\pi^+\pi^- $ is
calculated using formula:
\begin{gather}
\sigma^0_{\pi\pi} =\frac{N_{\pi\pi}}{IL\varepsilon_{\pi\pi} (1+\delta_r)}.
\end{gather}
Here $ 1 +\delta_r $ is a radiative correction, $ N_{\pi\pi} $ and
$\varepsilon_{\pi\pi} $ are the number of events and the detection efficiency for 
the process $ e^+ e^-\to\pi^+\pi^- $.
\end{enumerate}

The detection efficiency for each process is derived from the 
Monte Carlo simulation based on GEANT4 \cite{geant4, geant4-2}. Apparatus effects such as 
electronics noise, signal pile-up, actual time and amplitude
resolutions of electronics channels, the bad channels are 
taken into account in the simulation.

Generation of $e^+e^-\to e^+e^-$, $\mu^+\mu^-$ and $\pi^+\pi^-$ events is 
performed by the MCGPJ \cite{mcgpj} generator. It is based
on formulae from \cite{arbuzqed, arbuzhad}. The generator takes into
account initial and final state radiation (ISR and FSR), as well as
Coulomb interaction in the final state. It allows one to 
calculate cross sections and radiative corrections with accuracy
$\sigma_{rad}=0.2$\%.
The simulation of the process $e^+ e^-\to e^+ e^-$ is performed 
with the cut on the polar angles
of the final electron and positron $30^\circ<\theta_{e^\pm} <150^\circ$.

\begin{figure}[tbp]
\centering
\includegraphics[width=0.9\textwidth]{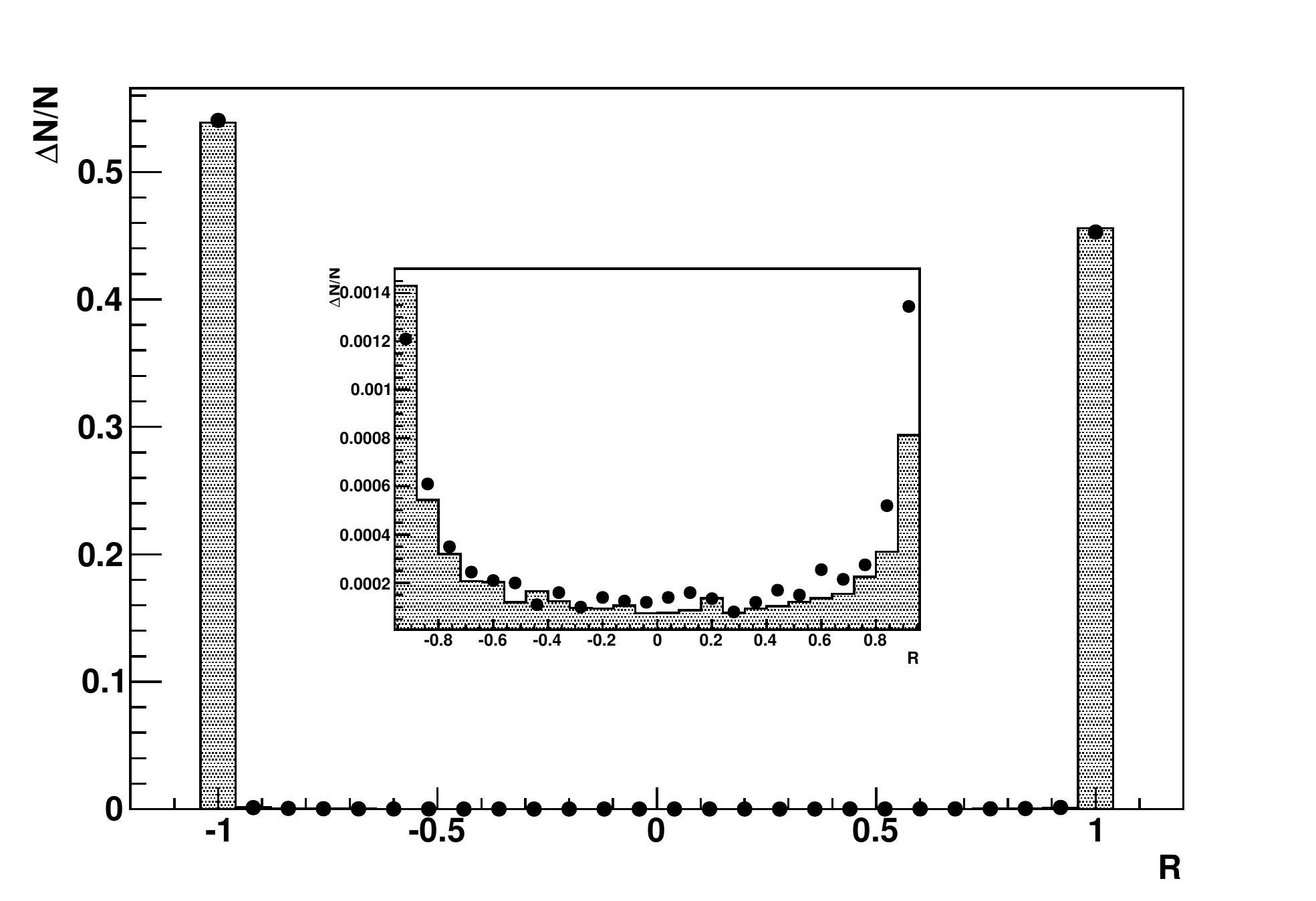}
\caption{\small The distribution of the separation parameter R for all collinear
events ($e^+e^- \to e^+e^-$, $\pi^+\pi^-$ and $\mu^+\mu^-$) at the energy
$\sqrt{s}=$778 MeV. The insert depicts the same histograms in the region between
the peaks. Dots -- experiment, histogram -- simulation.  Histogram for
MC simulation is sum of distributions for $e^+e^- \to \mu^+\mu^-$, $e^+e^-$ and
$\pi^+\pi^-$ events. The contribution of each process to the histogram was
calculated according to cross sections used in MCGPJ generator \cite{mcgpj}.}
\label{bdt}
\end{figure}

The $e^+e^-\to e^+e^-$, $\mu^+\mu^-$ and $\pi^+\pi^-$ events have
different distributions of the energy deposition over calorimeter crystals. In $e^+ e^-\to e^+ e^-$ 
events the electrons and positrons produce electromagnetic showers, with the 
most probable energy losses of about 0.92 of the initial particle energy. 
Muons lose their energy by ionization of the calorimeter material through
which they pass. The charged pions lose energy due to ionization and nuclear 
interaction with the detector material. 
The separation parameter of $e^+e^-\to e^+e^-$ and $e^+e^-\to\pi^+\pi^-$ 
events (R) in the energy region $\sqrt{s}=$ 0.5 -- 1.0 GeV is based on the differences 
in the energy deposition profiles. It was developed using machine learning method 
\cite{epi}. The distribution of the separation parameter $R$ is shown in 
figure~\ref{bdt}. The $e^+e^-\to e^+e^-$ events are located in
the region $ R <0$, while $e^+e^-\to\pi^+\pi^-,\mu^+\mu^- $ events are
located at $R>0$.

\subsection{Events selection}

During the data taking, the first-level trigger selects events with
one or more tracks in the drift chamber and with the total energy 
deposition in the calorimeter greater than 100 MeV. During processing of
the experimental data, event reconstruction is performed \cite{snd}.
For the further analysis the collinear events are selected using the following
criteria.
\begin{enumerate}
\item
The number of charged particles $ N_{cha}\ge 2 $. An event can also 
contain additional neutral particles due to beam background, 
nuclear interaction of charged pions, splitting of 
electromagnetic showers and initial and final state
radiation.
\item
$|\Delta\theta|=|180^\circ - (\theta_1 +\theta_2)|<12^\circ $ and
$|\Delta\phi|=|180^\circ-|\phi_1-\phi_2||<4^\circ $, where $\theta_{1,2}$ and
$\phi_{1,2}$ are the polar and azimuthal angles of charged particles with the 
largest energy deposition (particles in the event are ordered by the energy 
deposition), respectively.
\item
$E_{1,2}>40$ MeV, where $E_i$ is the energy deposition of the $i$th charged
particle.
\item
$ 50^\circ <\theta_0 <130^\circ$, where 
$\theta_0 = (\theta_1-\theta_2+180^\circ)/2$.
\item
$|r_{1,2}|<1 $ cm, where $r_i$ is the distance between the track of the $i$th
particle and the beam axis.
\item
$|z_{1,2}|<8 $ cm, where $z_i$ is the coordinate of the $i$th particle vertex (point of the track closest to the beam axis) 
along the beams axis.
\item
The muon system veto is used for suppressing the cosmic background.
\end{enumerate}

\subsection{Subtraction of $e^+e^-\to\pi^+\pi^-\pi^0 $ and cosmic background}

In the event sample selected under these conditions, one has 
$e^+ e^-\to e^+ e^-$, $\pi^+\pi^-$, $\mu^+\mu^-$ events, residual cosmic background, and a small contribution from $e^+ e^-\to\pi^+\pi^-\pi^0$ 
reaction at $\sqrt{s}\approx m_\omega$.

The number of background events from the process
$e^+e^-\to\pi^+\pi^-\pi^0$ is estimated as
\begin{gather}
N_{3\pi} = n_{3\pi}\times\frac{M_{3\pi}}{m_{3\pi}},
\end{gather}
where $ M_{3\pi} $ is a number of simulated $e^+e^-\to\pi^+\pi^-\pi^0$ events
selected using the nominal conditions for collinear events, described above, $n_{3\pi}$ and $m_{3\pi}$
are the number of data and simulated events, respectively, 
selected under conditions:
\begin{enumerate}
\item
$ N_{cha}\ge 2 $.
\item
The number of neutral particles $N_{neu}\ge 2$.
\item
$|\Delta\theta|>10^\circ$ and $|\Delta\phi|>10^\circ$.
\item
$ 40^\circ<\theta_{1,2}<140^\circ $.
\item
$\chi^2_{3\pi}<30$, where $\chi^2_{3\pi}$ is the $\chi^2$ of the kinematic 
fit of the event under $e^+e^-\to\pi^+\pi^-\pi^0$ hypothesis.
\end{enumerate}
It is found that the $e^+e^-\to\pi^+\pi^-\pi^0$ background is maximal in the energy point $\sqrt{s}=782.9$ MeV, where its fraction is less than 0.15\%, corresponding to 37 background events. 
\begin{figure} [tbp]
\centering
\includegraphics [width = 0.8\textwidth]{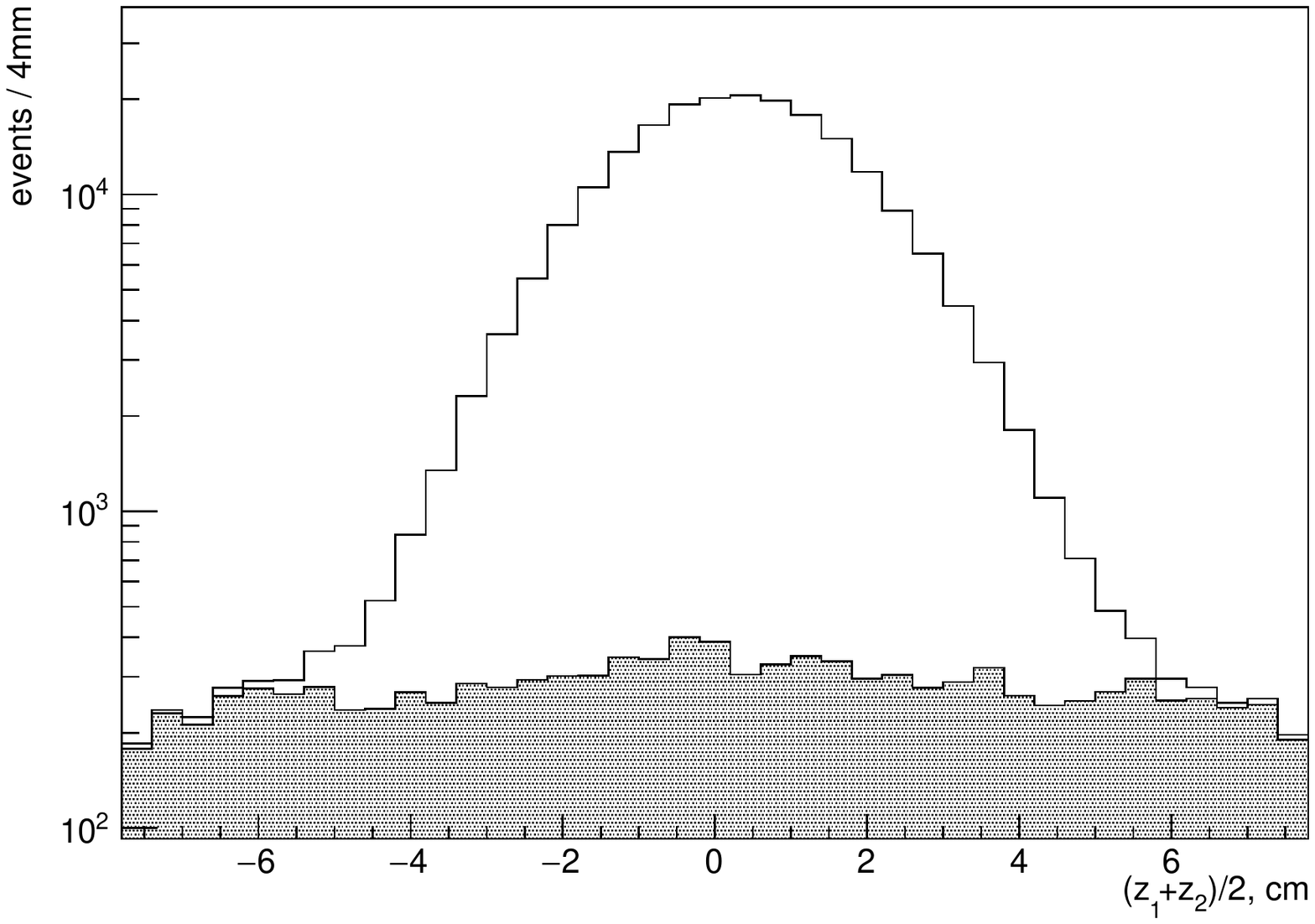}
\caption{The distributions of the $z$ coordinate of the charged particle vertex 
for collinear events at $\sqrt{s}=778$ MeV. The histogram represents events without muon 
system veto ($veto=0$), while the shaded histogram shows events with muon system veto.}
\label{z0}
\end{figure}

The cosmic events are suppressed by the muon system.
The $z$ coordinate distribution of the production point for collinear
events is shown in figure~\ref{z0}. The $e^+e^-$ annihilation events
have a Gaussian distribution peaked at $z=0$, while the cosmic distribution is nearly uniform. 
As figure~\ref{z0} shows, the muon subsystem veto ($veto=1$) separates cosmic 
muons from the $e^+e^-$ annihilation events.

The number of the residual cosmic events is estimated as 
follows
\begin{gather}
N_{cosm} = N_{data}^{veto = 1}\frac{N_{cosm}^{veto = 0}}{N_{cosm}^{veto = 1}},
\end{gather}
where $ N_{data}^{veto = 1} $ is the number of collinear events selected
using the nominal selection criteria, but with $veto=1$, 
$N_{cosm}^{veto=0}$ and $N_{cosm}^{veto=1}$ are the numbers of cosmic events 
with $veto=1$ and $veto=0$, respectively. Two types of cosmic events are used:
\begin{enumerate}
\item
Collinear events with additional cuts: $|r_{1,2}|> 0.5 $ cm and 
$|z_{1,2}|> 5 $ cm.
\item
Events recorded in special cosmic runs satisfying the nominal selection
criteria.
\end{enumerate}
In both cases, the ratio $N_{cosm}^{veto=0}/N_{cosm}^{veto=1}$ is found to be equal to 2.5\%$\pm$0.1\%.

\subsection{Detection efficiency}

\begin{figure}
\centering
\includegraphics [width = 0.8\textwidth]{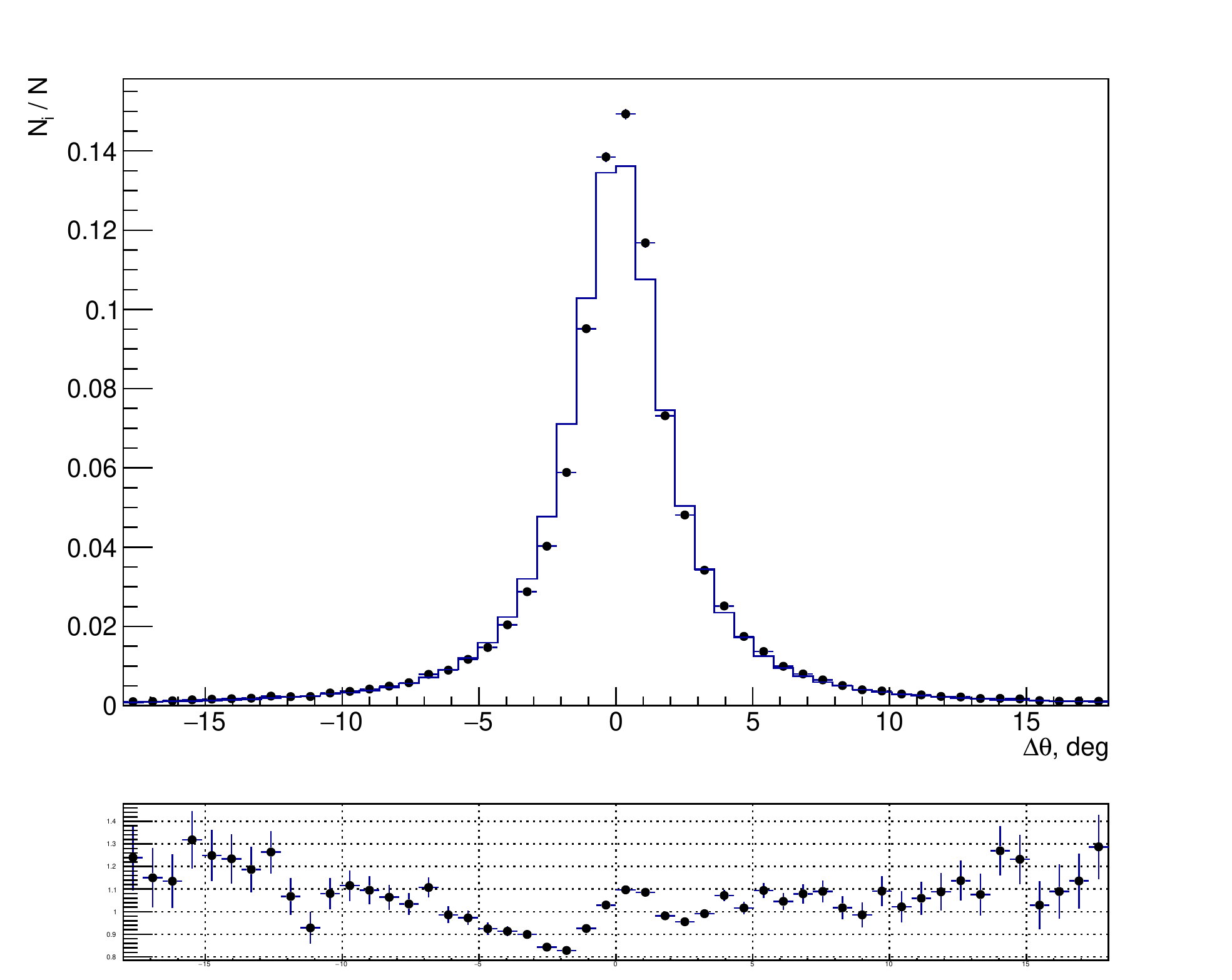}
\caption{\small The $\Delta\theta $ distribution for $e^+e^-\to\pi^+\pi^-$ 
events at $\sqrt{s} = 778 $ MeV. The solid histogram represents simulation, while the dotted histogram shows data. Their ratio depicted below.}
\label{D_th_ppc}
\end{figure}
\begin{figure}
\centering
\includegraphics [width = 0.8\textwidth]{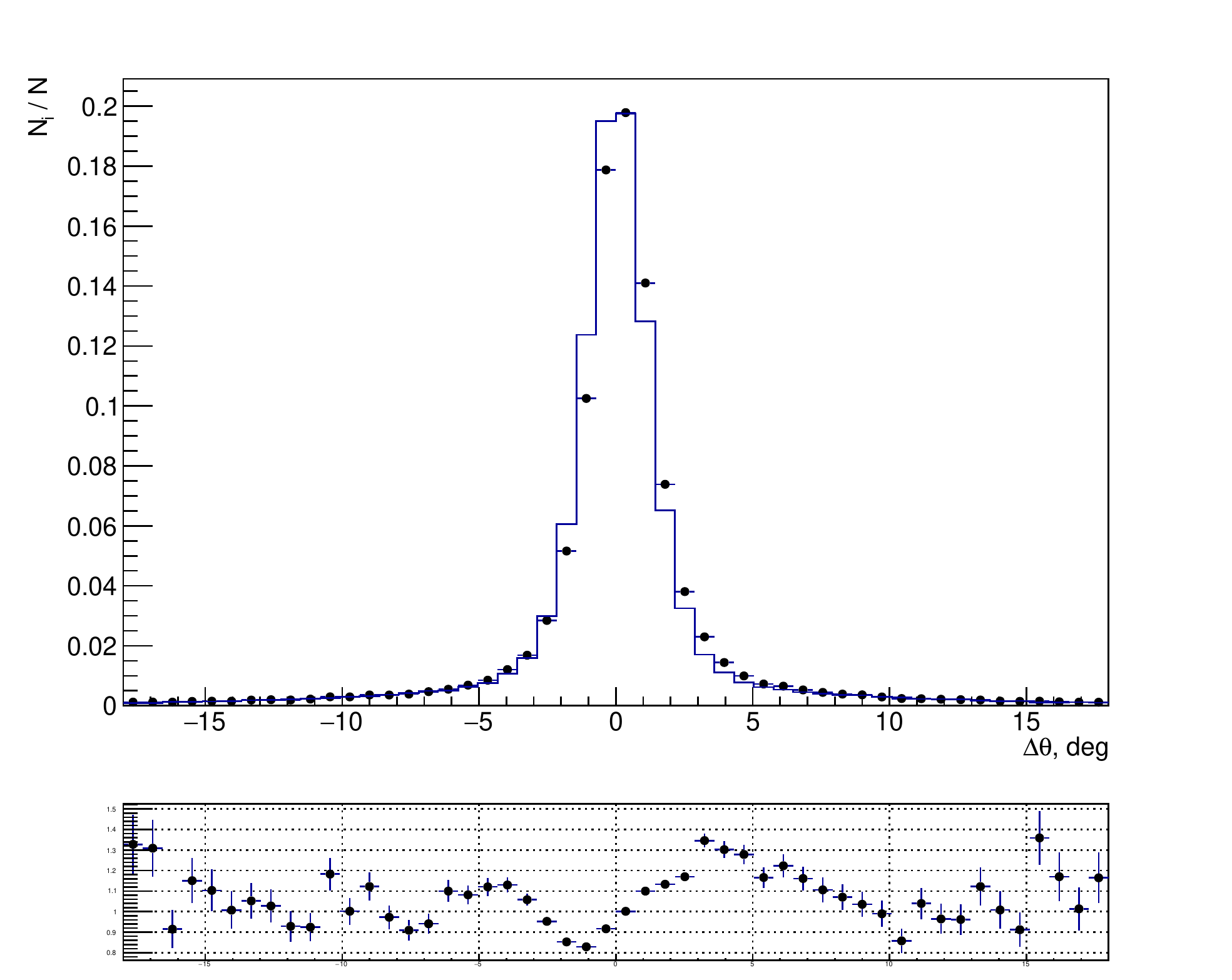}
\caption{\small The $\Delta\theta $ distribution for $e^+e^-\to e^+e^-$ 
events at $\sqrt{s} = 778 $ MeV. The solid histogram represents simulation, while the dotted histogram shows data. Their ratio depicted below.}
\label{D_th_ee}
\end{figure}
\begin{figure}
\centering
\includegraphics [width = 0.8\textwidth]{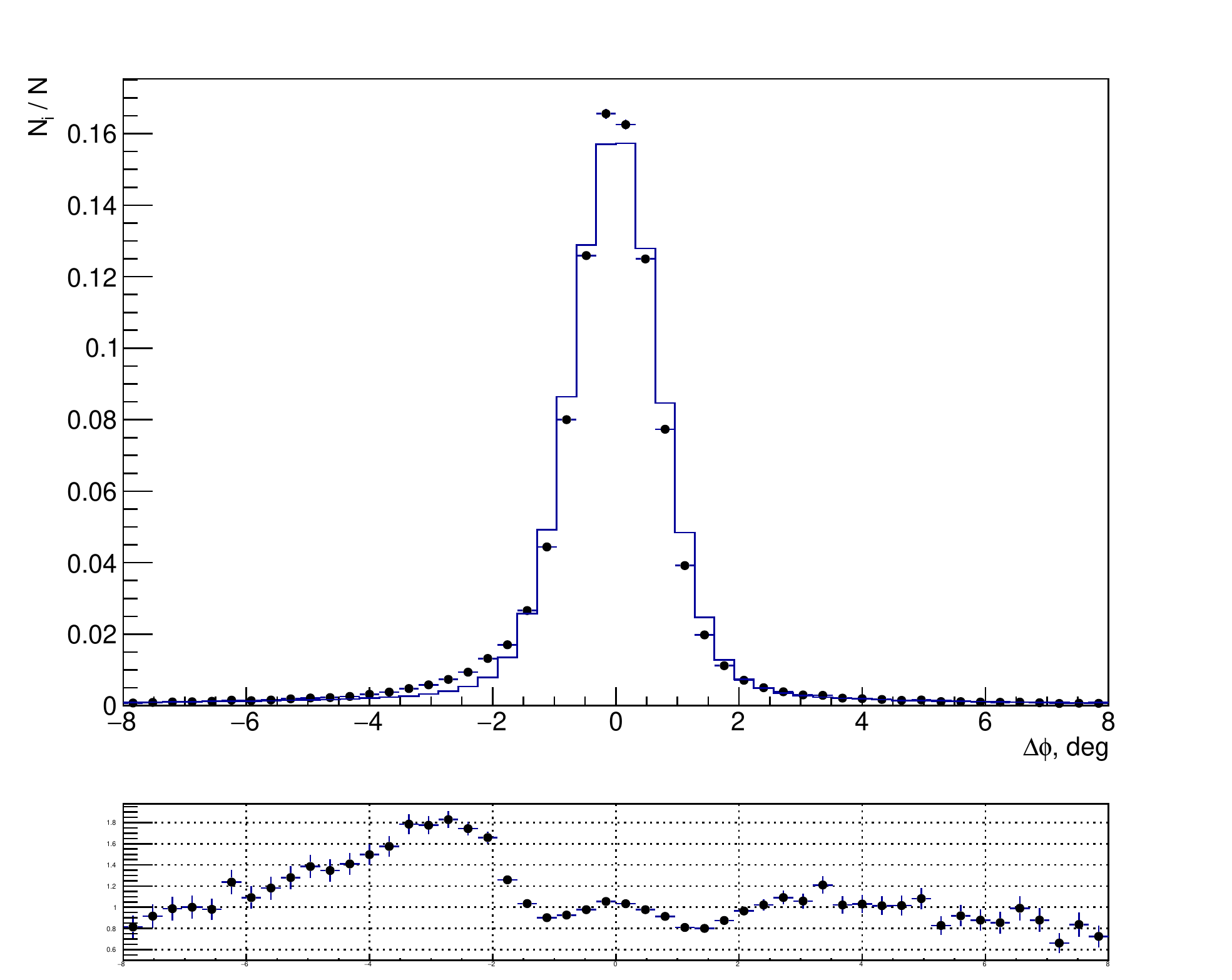}
\caption{\small The $\Delta\phi$ distribution for $e^+ e^-\to\pi^+\pi^-$ events at
$\sqrt{s}=778$ MeV. The solid histogram represents simulation, while the dotted histogram shows data. Their ratio depicted below.}
\label{D_phi_ppc}
\end{figure}
\begin{figure}
\centering
\includegraphics [width = 0.8\textwidth]{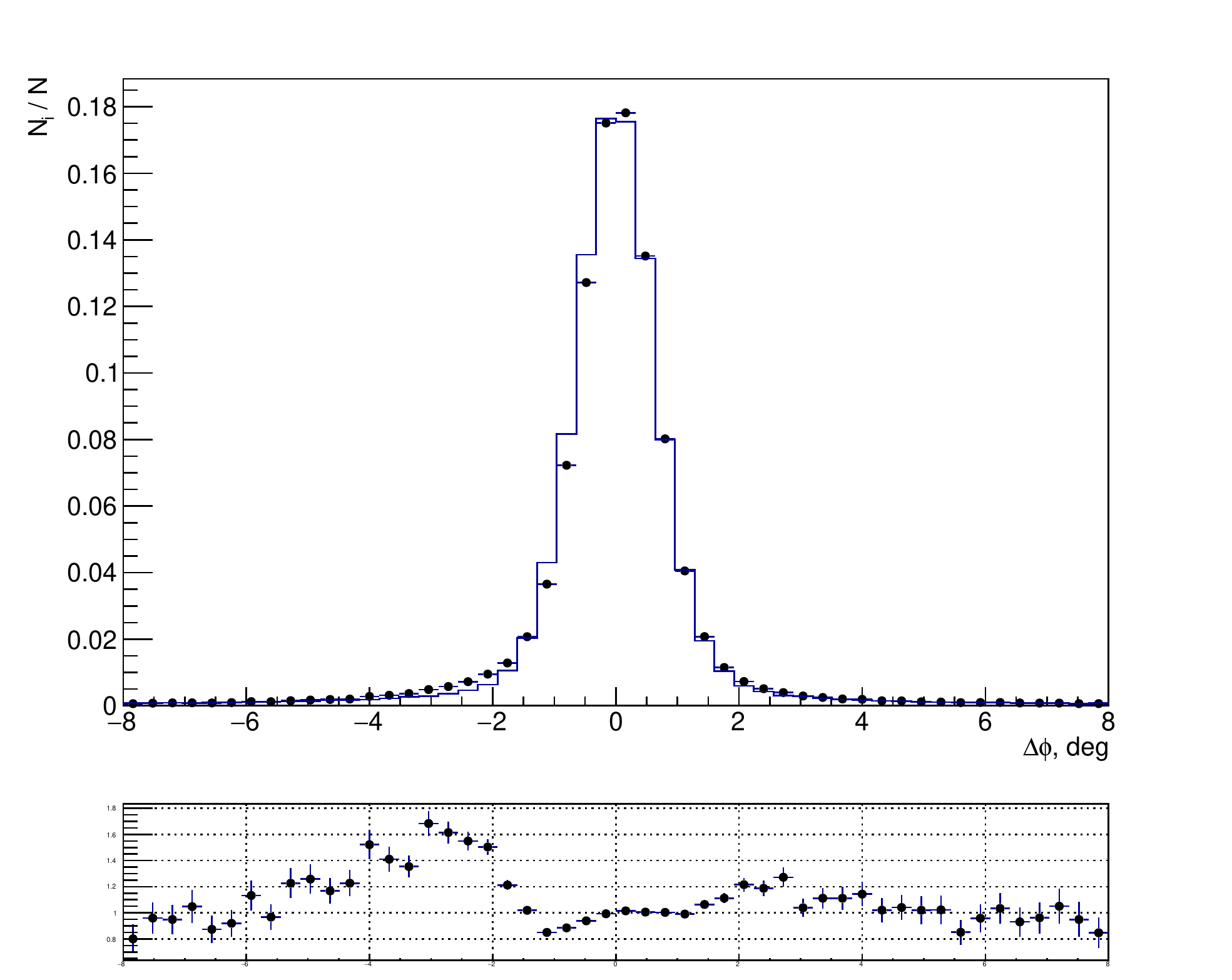}
\caption{\small The $\Delta\phi$ distribution for $e^+ e^-\to e^+e^-$ events at
$\sqrt{s}=778$ MeV. The solid histogram represents simulation, while the dotted histogram shows data. Their ratio depicted below.}
\label{D_phi_ee}
\end{figure}

The $\Delta\phi$ and $\Delta\theta$ distributions for the $e^+e^-\to e^+e^-$ 
and $e^+e^-\to\pi^+\pi^-$ events are shown in 
figure~\ref{D_th_ppc}, \ref{D_th_ee}, \ref{D_phi_ppc} and \ref{D_phi_ee}. There are
small differences in the shapes of the data and simulated spectra. 
The following values are used as a measure of the systematic uncertainty due to the $\Delta\theta$ and 
$\Delta\phi$ cuts:
\begin{center}
\begin{gather}
\delta_{x} =\frac{R_x^{\pi\pi}}{R_x^{ee}},\mbox{~~} x =\Delta\phi
(\Delta\theta).
\end{gather}
\end{center}
Here
\begin{align}
R_{\Delta\phi}^{i} &=\frac{N_i(|\Delta\phi|<4^\circ)}
{N_i(|\Delta\phi|<8^\circ)}/\frac{M_i(|\Delta\phi|<4^\circ)}
{M_i(|\Delta\phi|<8^\circ)},\\
R_{\Delta\theta}^{i} &= \frac{N_i(|\Delta\theta|<12^\circ)}
{N_i(|\Delta\theta|<18^\circ)}/\frac{M_i(|\Delta\theta|<12^\circ)}
{M_i (|\Delta\theta|<18^\circ)},
\end{align}
where $i=\pi\pi(ee)$, $N_i$ and $M_i$ are the numbers 
of data and simulated events selected under the conditions 
on $\Delta\phi $ and $\Delta\theta $ indicated in parentheses. 
The $\delta_{\Delta\theta}$ and $\delta_{\Delta\phi} $ do not depend
on energy. Their deviations from unity are taken as
systematic errors. Thus the systematic uncertainty associated with the $\Delta\phi $ and $\Delta\theta $ cuts
is $\sigma_\Delta=0.001\oplus 0.002\approx 0.002$.

\begin{figure}
\centering
\includegraphics [width = 0.8\textwidth]{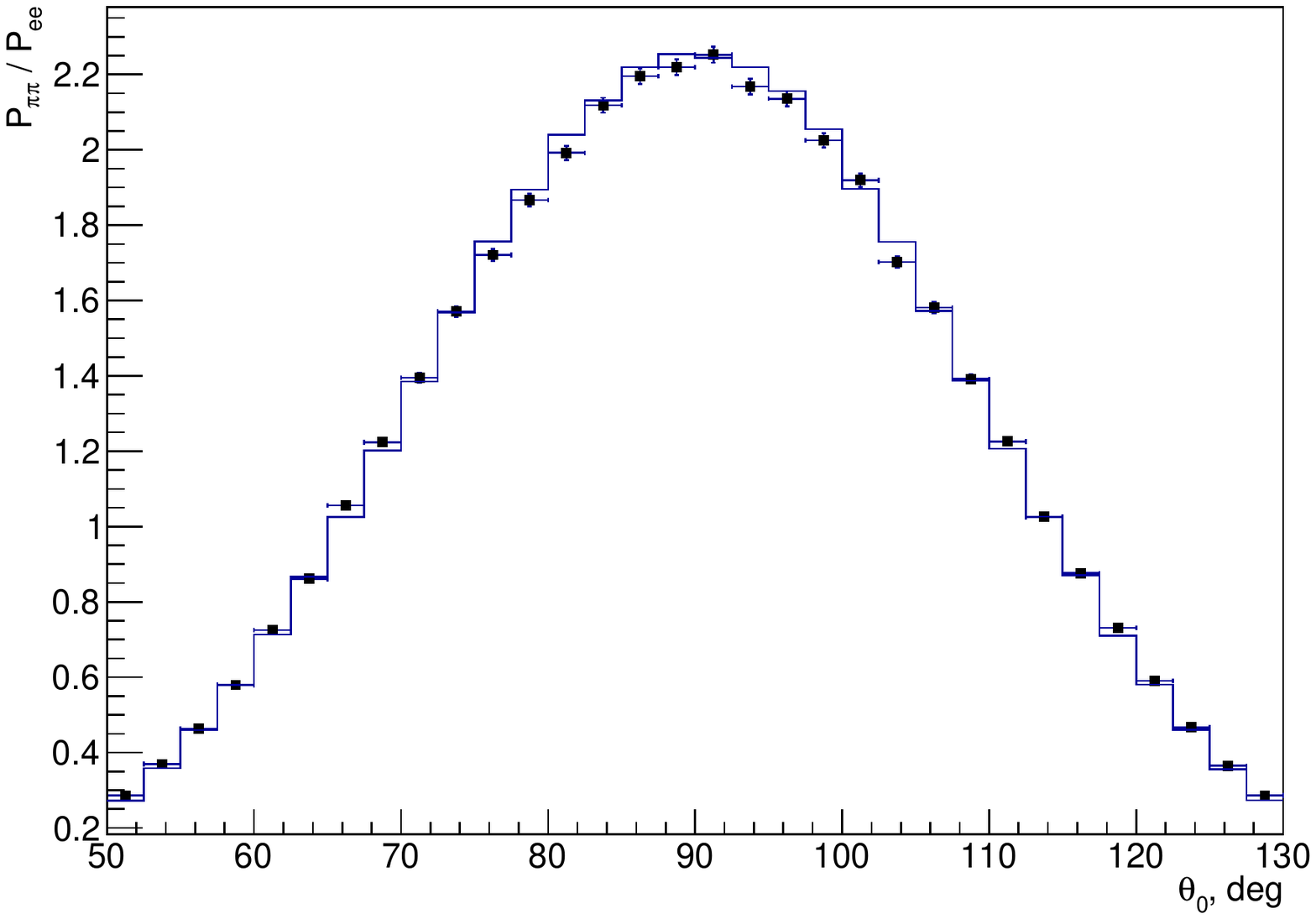}
\caption{\small The ratio of $\theta_0$ distributions of 
$e^+ e^-\to\pi^+\pi^-$ and $e^+ e^-\to e^+ e^-$ events. 
Histogram -- simulation, dots -- experiment.}
\label{th0}
\end{figure}

\begin{figure}
\centering
\includegraphics [width = 0.8\textwidth]{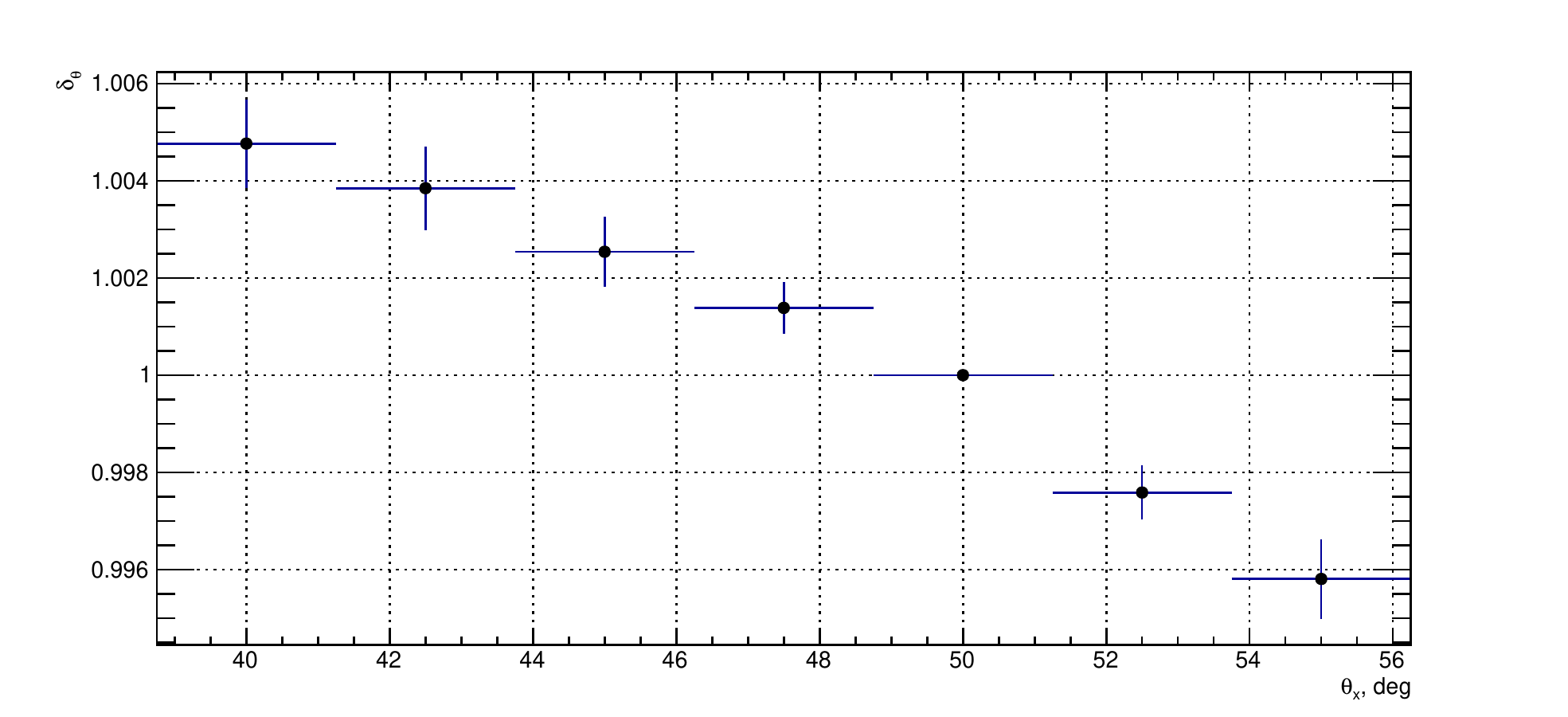}
\caption{\small The $\delta_\theta$ dependence on $\theta_x$
(\ref{tet0}).}
\label{del_th0}
\end{figure}

The ratio of the $\theta_{0}$ distributions for the $e^+e^-\to\pi^+\pi^-$ and 
$e^+e^-\to e^+e^-$ events is shown in figure~\ref{th0}. There are some 
differences between these ratios for data and simulated distributions.
To estimate the systematic error due to the $\theta_0$ cut, the following ratio is used:
\begin{gather}
\delta_\theta={\delta(\theta_x)\over\delta(50^\circ)}, \mbox{~~} 
40^\circ<\theta_x<55^\circ,
\end{gather}
where
\begin{gather}
\label{tet0}
\delta(\theta_x)=
\frac{N_{\pi\pi}(\theta_x<\theta<180^\circ-\theta_x)}
{N_{ee}(\theta_x<\theta<180^\circ-\theta_x)} /
\frac{M_{\pi\pi}(\theta_x<\theta<180^\circ-\theta_x)}
{M_{ee}(\theta_x<\theta<180^\circ-\theta_x)}.
\end{gather}
Here $ N_{\pi\pi} (\theta_x <\theta_0 <180^\circ-\theta_x) $,
$ N_{ee} (\theta_x <\theta_0 <180^\circ-\theta_x) $,
$ M_{\pi\pi} (\theta_x <\theta_0 <180^\circ-\theta_x) $,
$ M_{ee} (\theta_x <\theta_0 <180^\circ-\theta_x) $ are
the numbers of $ e^+ e^-\to\pi^+\pi^- $ and $ e^+ e^-\to e^+ e^- $ events 
summed over all energy points in experiment and simulation with 
$\theta_x <\theta_0 <180^\circ-\theta_x $. The largest deviation of 
$\delta_{\theta_0}$ from unity is equal to 0.005 (figure~\ref{del_th0}). 
This value is taken as a systematic error $\sigma_\theta$ associated with 
$50^\circ<\theta_0<130^\circ $ cut.


Imperfection in simulation of pion nuclear interactions
implies that the cut on the particle energy deposition
leads to an inaccuracy in the detection efficiency of
the $e^+e^-\to\pi^+\pi^-$ process. To take this inaccuracy into
account, the detection efficiency is multiplied by the
correction coefficient. The correction coefficient
is obtained by using pseudo $\pi\pi$ events, which are constructed
using events of the processes $e^+e^-\to\omega(\phi)\to\pi^+\pi^-\pi^0$ 
and $e^+e^-\to\pi^+\pi^-$ \cite{epi}. The corrections obtained using
different types of pseudo events differ less than 0.005, they do not
depend on the pion energy and their average is equal to 0.992. 
As a result, the correction coefficient is set equal to 0.992, and the
difference is taken as a systematic error $\sigma_E=0.005$.

\begin{figure}
\centering
\includegraphics [width = 0.8\textwidth]{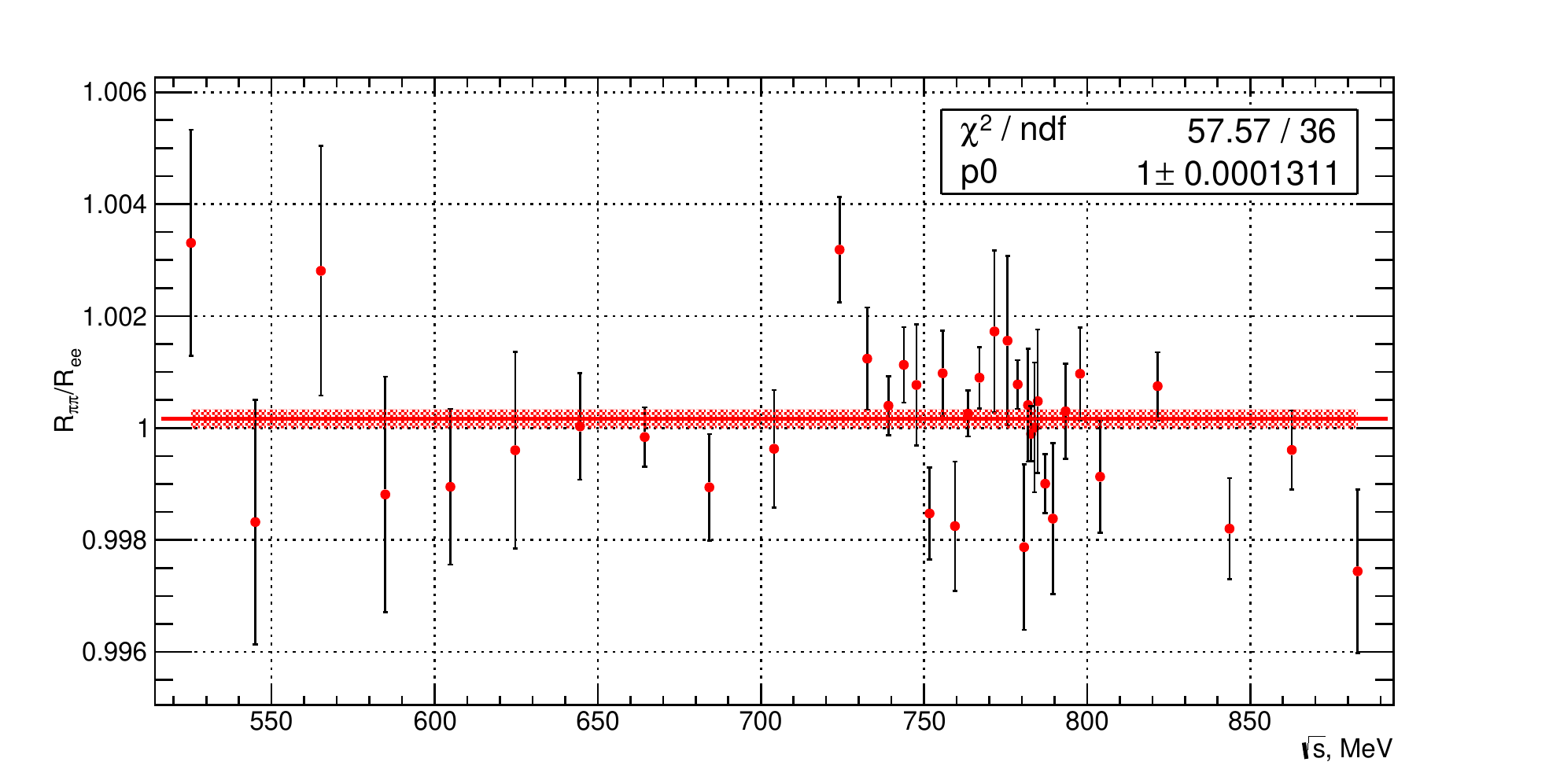}
\caption{\small The $R_{\pi\pi}/R_{ee}$ dependence on $\sqrt{s}$. Line depicts an average value.}
\label{fig:trk}
\end{figure}

In the tracking system, the particle track can be lost due to reconstruction 
inefficiency. The probabilities $\varepsilon^{data}_{\pi\pi}$ and
$\varepsilon^{data}_{ee}$ to find two tracks in the $e^+e^-\to\pi^+\pi^-$ 
and $e^+ e^-\to e^+ e^-$ events are determined using experimental data.
Their ratio to probabilities derived from simulated events
\begin{gather}
R_{i} =\frac{\varepsilon^{data}_{j}}{\varepsilon^{mc}_{j}},
\mbox{~~} i = ee,\pi\pi
\end{gather}
can vary significantly in the different energy points. But the ratio
$R_{\pi\pi}/R_{ee}$, which contributes to the measured cross section, is 
energy independent and equal to unity with error $10^{-4}$ (figure~\ref{fig:trk}).

Pions can be lost due to the nuclear interaction in the
detector material before the tracking system.
The probability of pion loss is studied using $e^+e^-\to\pi^+\pi^-\pi^0$ events. It was found that 
the difference between these values in data and simulation is
0.002, which is taken as a systematic error $\sigma_{nucl}=0.2$\%. 

The use of the muon system veto for event selection ($veto=0$) leads to 
inaccuracy in the determination of the measured cross section due to the 
uncertainty in the simulation of the muons and pions traversing the detector.
To obtain the necessary corrections, the events close to the median plane
$0^\circ<\phi<14^\circ $, $166^\circ<\phi<194^\circ$, $360^\circ>\phi>346^\circ$), where
the cosmic background is minimal, are used.
The correction is the ratio of the $e^+e^-\to\pi^+\pi^-$ cross sections 
measured with ($veto=0$) and without ($veto\ge 0$) using the muon system:

\begin{gather}
\delta_{veto} =\frac{\sigma_{\pi\pi} (veto\ge 0)}
{\sigma_{\pi\pi}(veto = 0)}.
\end{gather}

\begin{figure}
\centering
\includegraphics [width = 0.8\textwidth]{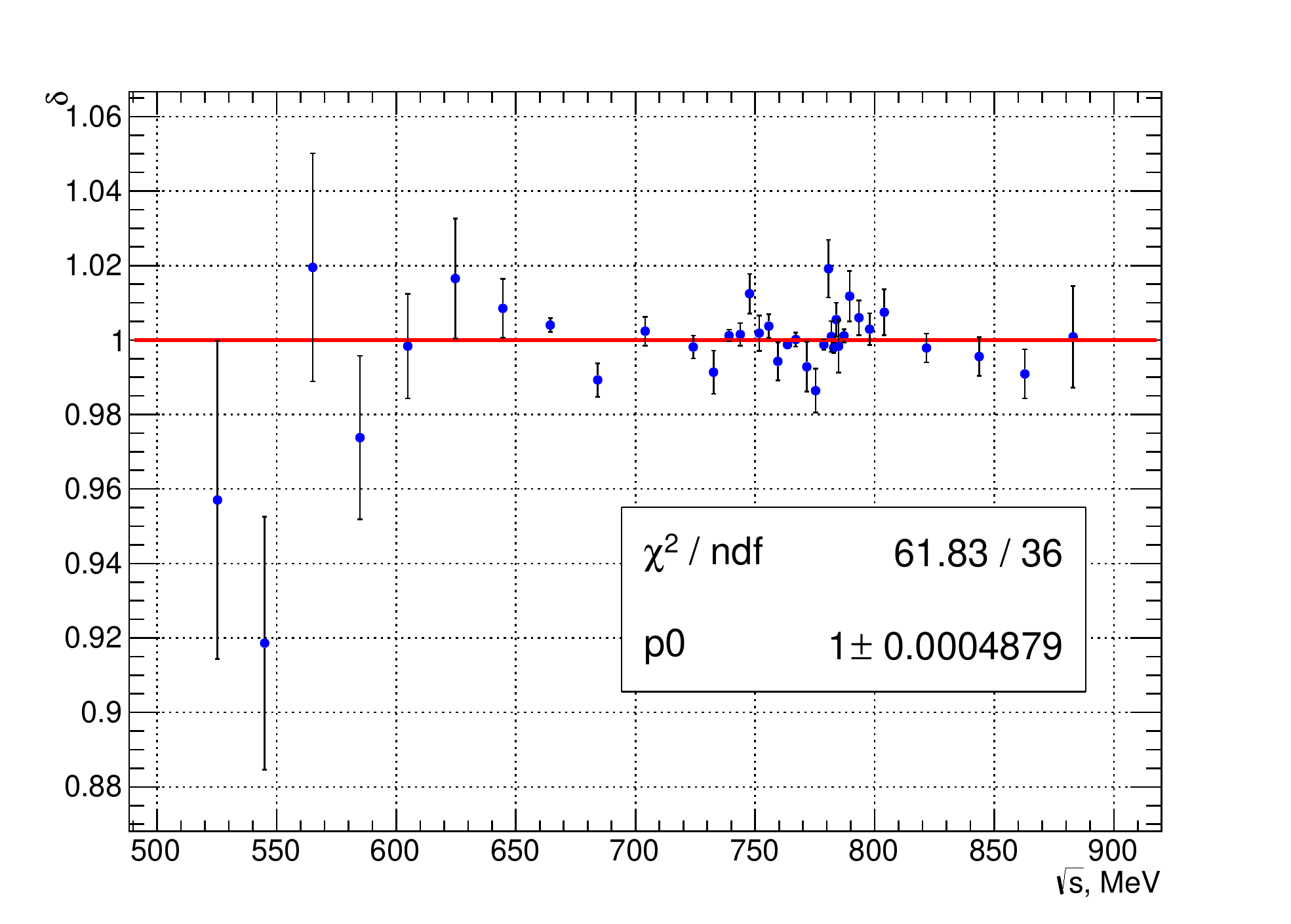}
\caption{\small The $\delta_{veto}$ dependence on $\sqrt{s}$. Line depicts an average value.}
\label{fig:veto}
\end{figure}

In the case of $ veto\ge 0 $, a contribution of the residual cosmic muons 
background is estimated from the fit to the $(z_1+z_2)/2$ spectrum with a sum
of the Gaussian and uniform distributions. The $\delta_{veto}$ does
not dependent on energy and its average value is consistent with 1 (figure~\ref{fig:veto}). 
This indicates the absence of the systematic error related to the condition $veto=0$.
Relatively high $\chi^2/n.d.f$ in figures~\ref{fig:veto} (1.71) and ~\ref{fig:trk} (1.6) is due to the large devitions of 2--3 energy points. It's caused by background contamination of the control samples (events with $veto\ge 0$ or only one reconstructed track) used in $\delta_{veto}$ and $R_{\pi\pi}/R_{ee}$ calculations. 


Trigger efficiency is greater than 99.9\% for all types of collinear
events due to the energy deposition cuts $E_{1,2} >40$ MeV. These cuts
provide performance of the energy deposition threshold. Therefore
systematic uncertainty from trigger inefficiency is considered to be
negligible.

Uncertainties in simulation of energy depositions in the calorimeter can lead to an inaccuracy in e/$\pi$ discrimination.  
The identification efficiency and related systematic error were
studied in \cite{epi} using pseudo--$\pi\pi$ and pseudo--$ee$ events. It varies with energy from 0.996 to 0.998
for the $e^+e^- \to e^+e^-$ events and from 0.994 to 0.998 for the $e^+e^-
\to \pi^+\pi^-$ events. The systematic error $\sigma_{PID}$ of the $e^+e^-\to\pi^+\pi^-$
cross section measurement due to cuts R$<0.0$ and R$>0.0$ does not exceed
0.002 at $\sqrt{s}>$650 MeV. Below 650 MeV, the $\sigma_{PID}$ value
increases with decrease of energy and reaches 0.005 at $\sqrt{s}$=525.1 MeV.

\subsection{Calculation of the cross section}

The number of selected events in the regions $R>$0 and $R<$0 are:
\begin{equation}
N_a = IL (\sigma_{\pi\pi}\varepsilon^a_{\pi\pi}+
\sigma_{\mu\mu}\varepsilon^a_{\mu\mu}+
\sigma_{ee}\varepsilon^a_{ee}) + N^a_{nc},
\end{equation}
where index $a=1,2$ indicates the events with 0$<R$ 
and $R>$0 respectively; $\sigma_{jj}$ and $\varepsilon^a_{jj} $ are physical 
cross section and detection efficiency of the process with 
jj = $\pi^+\pi^-,\mu^+\mu^-,e^+e^-$ in the final state, $N^a_{nc}$ is a 
number of $e^+e^-\to\pi^+\pi^-\pi^0$ and cosmic background events, $IL$ is the 
integrated luminosity. The detection efficiencies
$\varepsilon^a_{jj}$ take into account the correction coefficients described 
above. Using the formula for $N_a$, the $e^+e^-\to\pi^+\pi^-$ cross 
section is calculated as
\begin{equation}\label{CS:ppc}
\sigma_{\pi\pi} (s_i) =\frac{N_1-N^1_{nc} -
IL\sigma_{\mu\mu}\varepsilon^1_{\mu\mu} (s_i) -
\sigma_{ee}\varepsilon^1_{ee}}{IL\varepsilon^1_{\pi\pi}}, 
\end{equation}
where
\begin{equation}\label{lumi}
IL =\frac{(N_2-N^2_{nc})\varepsilon^1_{\pi\pi} - 
(N_1-N^1_{nc})\varepsilon^2_{\pi\pi}}{\sigma_{ee} 
(\varepsilon^2_{ee}\varepsilon^1_{\pi\pi} -
\varepsilon^1_{ee}\varepsilon^2_{\pi\pi}) +
\sigma_{\mu\mu} (\varepsilon^2_{\mu\mu}\varepsilon^1_{\pi\pi} -
\varepsilon^1_{\mu\mu}\varepsilon^2_{\pi\pi})}.
\end{equation}
 
Subtraction of the $e^+ e^-\to\mu^+\mu^-$ background leads to additional 
contribution to the systematic error, which is estimated as follows:
\begin{equation}\label{mu_sys}\centering
\sigma_{\mu} = (\sigma_{\theta}\oplus\sigma_{\Delta}\oplus\sigma_{rad})
\times\frac{\varepsilon^1_{\mu\mu}\sigma_{\mu\mu}}
{\varepsilon^1_{\pi\pi}\sigma_{\pi\pi}}.
\end{equation}

The Born cross section for the $e^+e^-\to\pi^+\pi^-$ process is
calculated from as
\begin{equation}\label{CS:born}
\sigma^0_{\pi\pi} (s_i) =\frac{\sigma_{\pi\pi} (s_i)}{1+\delta_{rad} (s_i)}.
\end{equation}
The radiative correction $\delta_{rad}(s_i)$, which takes into account the 
initial and final states radiation, is calculated using the MCGPJ generator. 
The value of $\delta_{rad}(s)$ depends on the $\sigma^0_{\pi\pi}(s)$ cross 
section at lower energies, and it is therefore calculated iteratively.
The iteration stops when its value changes by less than 0.05\% in 
consecutive iterations. The correction for the center of mass energy
spread is taken into account also. The spread does not exceed 0.3 MeV
in the energy region below 1 GeV, and the correction is less than 0.1\%.

The measured cross section $\sigma^0_{\pi\pi}$ is presented in 
table~\ref{tab:pi2tab1}. The systematic errors of the cross section 
determination are listed in table~\ref{tab:sys}.

\begin{table}
\begin{center}
\begin{tabular}{|c|c|c|c|c|}\hline
$\sqrt{s}$, MeV&$\sigma_{\pi\pi}$, nb&
$\sigma^{0}_{\pi\pi}$, nb& $|F(s)|^{2}$&$\sigma_{bare}$, nb\\ \hline
525.1&203.4$\pm$12.3$\pm$2.4&210.4$\pm$12.7$\pm$2.5&4.4$\pm$0.3$\pm$0.1&209.7$\pm$12.7$\pm$2.5 \\
544&224.4$\pm$10.1$\pm$2.5&232.5$\pm$10.5$\pm$2.6&5$\pm$0.2$\pm$0.1&231.9$\pm$10.4$\pm$2.6 \\
565.2&235$\pm$12.3$\pm$2.4&244.3$\pm$12.8$\pm$2.5&5.5$\pm$0.3$\pm$0.1&243.8$\pm$12.8$\pm$2.5 \\
585&254.2$\pm$10.7$\pm$2.5&265$\pm$11.1$\pm$2.6&6.2$\pm$0.3$\pm$0.1&264.8$\pm$11.1$\pm$2.6 \\
604.8&328.8$\pm$8.7$\pm$3&344.7$\pm$9.2$\pm$3.1&8.3$\pm$0.2$\pm$0.1&344.8$\pm$9.2$\pm$3.1 \\
624.8&366.4$\pm$11.1$\pm$3.2&386.1$\pm$11.7$\pm$3.4&9.7$\pm$0.3$\pm$0.1&386.7$\pm$11.7$\pm$3.4 \\
644.6&438$\pm$8.2$\pm$3.7&464.2$\pm$8.7$\pm$3.9&12.1$\pm$0.2$\pm$0.1&465.6$\pm$8.7$\pm$3.9 \\
664.5&525.9$\pm$3.5$\pm$4.4&561.3$\pm$3.7$\pm$4.7&15.3$\pm$0.1$\pm$0.1&563.7$\pm$3.7$\pm$4.7 \\
684.4&642.1$\pm$8.4$\pm$5.3&689.1$\pm$9$\pm$5.6&19.5$\pm$0.3$\pm$0.2&692.9$\pm$9.1$\pm$5.7 \\
704.2&798.1$\pm$10.3$\pm$6.5&860.7$\pm$11.1$\pm$7&25.4$\pm$0.3$\pm$0.2&865.5$\pm$11.1$\pm$7 \\
724.1&1030.4$\pm$9.5$\pm$8.3&1112.6$\pm$10.3$\pm$9&34.2$\pm$0.3$\pm$0.3&1116.6$\pm$10.3$\pm$9 \\
739.1&1146.4$\pm$5.6$\pm$9.2&1233.7$\pm$6$\pm$9.9&39.1$\pm$0.2$\pm$0.3&1234$\pm$6$\pm$9.9 \\
743.8&1200.9$\pm$9.8$\pm$9.7&1289.4$\pm$10.6$\pm$10.4&41.3$\pm$0.3$\pm$0.3&1288.1$\pm$10.6$\pm$10.4 \\
747.7&1215$\pm$14.4$\pm$9.8&1301.6$\pm$15.4$\pm$10.5&42$\pm$0.5$\pm$0.3&1298.7$\pm$15.4$\pm$10.5 \\
751.7&1199.4$\pm$13.7$\pm$9.7&1281.4$\pm$14.7$\pm$10.3&41.7$\pm$0.5$\pm$0.3&1276.6$\pm$14.6$\pm$10.3 \\
755.7&1246.5$\pm$10.8$\pm$10&1327.9$\pm$11.5$\pm$10.7&43.5$\pm$0.4$\pm$0.4&1321.3$\pm$11.4$\pm$10.6 \\
759.6&1288.3$\pm$17.3$\pm$10.4&1368$\pm$18.3$\pm$11&45.2$\pm$0.6$\pm$0.4&1360.3$\pm$18.2$\pm$10.9 \\
763.6&1263.4$\pm$5$\pm$10.2&1336.8$\pm$5.2$\pm$10.8&44.5$\pm$0.2$\pm$0.4&1328.9$\pm$5.2$\pm$10.7 \\
767.8&1249.1$\pm$6.9$\pm$10.1&1317$\pm$7.2$\pm$10.6&44.2$\pm$0.2$\pm$0.4&1310$\pm$7.2$\pm$10.5 \\
771.6&1290.3$\pm$22.2$\pm$10.4&1356.5$\pm$23.3$\pm$10.9&45.9$\pm$0.8$\pm$0.4&1351.7$\pm$23.2$\pm$10.9 \\
775.7&1290.9$\pm$17.2$\pm$10.4&1353.6$\pm$18$\pm$10.9&46.2$\pm$0.6$\pm$0.4&1353.2$\pm$18$\pm$10.9 \\
778.6&1257$\pm$5.3$\pm$10.1&1311.1$\pm$5.5$\pm$10.5&45$\pm$0.2$\pm$0.4&1307.4$\pm$5.5$\pm$10.5 \\
780.7&1198.9$\pm$18.4$\pm$9.7&1229.2$\pm$18.9$\pm$9.9&42.3$\pm$0.7$\pm$0.3&1211.4$\pm$18.6$\pm$9.8 \\
782&1104.8$\pm$11.2$\pm$8.9&1106.9$\pm$11.2$\pm$8.9&38.2$\pm$0.4$\pm$0.3&1074.7$\pm$10.9$\pm$8.7 \\
782.9&1058.1$\pm$4.8$\pm$8.5&1039.8$\pm$4.7$\pm$8.4&36$\pm$0.2$\pm$0.3&999$\pm$4.5$\pm$8 \\
783.7&1004.9$\pm$11.6$\pm$8.1&971.9$\pm$11.3$\pm$7.8&33.7$\pm$0.4$\pm$0.3&925.2$\pm$10.7$\pm$7.5 \\
784.7&959.2$\pm$12.8$\pm$7.7&916.8$\pm$12.2$\pm$7.4&31.9$\pm$0.4$\pm$0.3&865.8$\pm$11.6$\pm$7 \\
786.7&913.5$\pm$5.1$\pm$7.4&872.3$\pm$4.8$\pm$7&30.4$\pm$0.2$\pm$0.2&819.1$\pm$4.5$\pm$6.6 \\
789.5&934.6$\pm$14.1$\pm$7.5&903.1$\pm$13.7$\pm$7.3&31.7$\pm$0.5$\pm$0.3&850.9$\pm$12.9$\pm$6.9 \\
793.9&890.4$\pm$10$\pm$7.2&867.8$\pm$9.7$\pm$7&30.7$\pm$0.3$\pm$0.2&823.1$\pm$9.2$\pm$6.6 \\
797.7&858.9$\pm$10.1$\pm$6.9&836.3$\pm$9.9$\pm$6.7&29.8$\pm$0.4$\pm$0.2&795.8$\pm$9.4$\pm$6.4 \\
804&819.5$\pm$10.5$\pm$6.6&791.4$\pm$10.1$\pm$6.4&28.6$\pm$0.4$\pm$0.2&755.4$\pm$9.6$\pm$6.1 \\
821.8&654.8$\pm$5.6$\pm$5.3&608.7$\pm$5.2$\pm$4.9&22.8$\pm$0.2$\pm$0.2&583$\pm$5$\pm$4.7 \\
843.4&496.6$\pm$5.8$\pm$4&438$\pm$5.1$\pm$3.6&17.1$\pm$0.2$\pm$0.1&420.4$\pm$4.9$\pm$3.4 \\
862.7&382.2$\pm$4.6$\pm$3.1&321.2$\pm$3.9$\pm$2.6&13$\pm$0.2$\pm$0.1&309$\pm$3.7$\pm$2.5 \\
883.2&303.2$\pm$6.7$\pm$2.5&242.1$\pm$5.3$\pm$2&10.2$\pm$0.2$\pm$0.1&233.5$\pm$5.1$\pm$1.9 \\
\hline    
\end{tabular} 
\caption{The results of the $e^+e^-\to\pi^+\pi^-$ cross section measurements.
$\sigma_{\pi\pi}$, $\sigma^0_{\pi\pi}$ and $F(s)$ are the physical, Born cross 
sections of the process $e^+ e^-\to\pi^+\pi^-$, and pion form factor, calculated with formula in ref.~\cite{snd2pi}. 
$\sigma_{bare}$ is the $e^+ e^-\to\pi^+\pi^-$ undressed cross section without
vacuum polarization, but with the final state radiative correction. Both statistical and systematic errors are shown. \bigskip}
\label{tab:pi2tab1}
\end{center} 
\end{table} 

\begin{table}
\begin{center}
\begin{tabular}{|c|c|c|}\hline
Error & at $\sqrt{s}> 600$ MeV, \% & at $\sqrt{s} \le 600 $ MeV, \% \\\hline
$\sigma_{PID}$ & 0.1--0.2 & 0.3--0.5 \\
$\sigma_\mu $ & 0.0--0.2 & 0.3--0.7 \\\hline
$\sigma_\Delta$ &\multicolumn{2}{c|}{0.2} \\
$\sigma_\theta$ &\multicolumn{2}{c|}{0.5} \\
$\sigma_E$ &\multicolumn{2}{c|}{0.5} \\
$\sigma_{rad}$ &\multicolumn{2}{c|}{0.2} \\
$\sigma_{nucl}$ &\multicolumn{2}{c|}{0.2} \\\hline
total & 0.8 & 0.9--1.2 \\\hline
\end{tabular}
\caption{Various contributions to the systematic error of the
$e^+e^-\to\pi^+\pi^-$ cross section measurement.
\bigskip}
\label{tab:sys}
\end{center}
\end{table}

\subsection{Fit to the measured cross section}

In the framework of the vector meson dominance model, the cross section of the 
$e^+e^-\to\pi^+\pi^-$ process is
\begin{equation}\label{CS:S}
\sigma^0_{\pi\pi} (s) ={2\over3}{\alpha^2\over s^{5/2}} P_{\pi\pi}(s)
|A_{\pi\pi} (s)|^2,
\end{equation}
where $P_{\pi\pi}(s)$ is the phase space factor:
\begin{equation}
P_{\pi\pi}(s)=q^3_{\pi}(s), \mbox{~~~} q_\pi (s)={1\over 2}\sqrt{s-4m^2_\pi}.
\end{equation}
The transition amplitudes are given by
\begin{equation}
|A_{\pi\pi} (s)|^2 =\Biggl|\sqrt{3\over 2}{1\over\alpha}
\sum_{V =\rho,\omega,\rho^\prime}
{{\Gamma_V m_V^3\mbox{~}\sqrt{m_V\sigma (V\to\pi^+\pi^-)}}\over{D_V (s)}}
{{e^{i\phi_{\rho V}}\over{\sqrt{q^3_{\pi}(m_V)}}}}\Biggr|^2,
\end{equation}
where 
\begin{align}
D_V(s)&=m_V^2-s-i\mbox{~}\sqrt{s}\Gamma_V(s),\\
\Gamma_V (s) &=\sum_{f}\Gamma (V\to f, s).
\end{align}
Here, $f$ denotes the final state of the vector meson $V$ decay, $m_V$ is the 
vector meson mass, $\Gamma_V=\Gamma_V(m_V)$ and $\phi_{\rho V}$ is the 
relative interference phase between the vector mesons V and $\rho$, 
and, hence, $\phi_{\rho\rho}=0$.

\begin{figure} [tbp]
\centering
\includegraphics [width = 0.8 \textwidth]{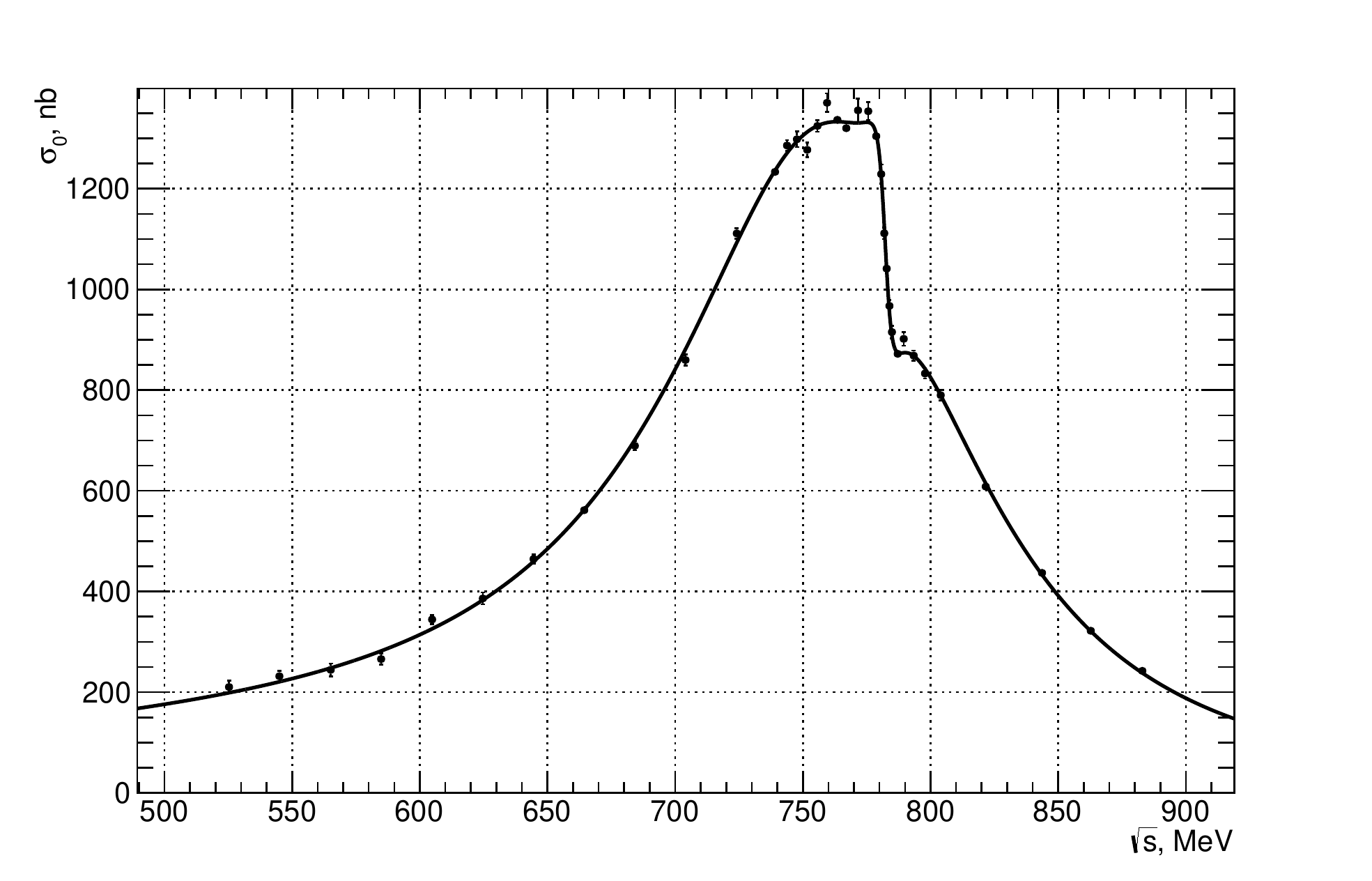}
\caption{The dependence of the Born cross section of the 
$e^+e^-\to\pi^+ \pi^-$ process on energy,
dots with errors are data, curve is the fit result.}
\label{fig:pi2fit}
\end{figure}

The following forms of the energy dependence of the
vector meson total widths are used:
\begin{align}
\Gamma_\omega (s) &={m_\omega^2\over s}{q^3_\pi (s)\over q^3_\pi (m_\omega)}
\Gamma_\omega B_{\omega\to\pi^+\pi^-} +
{q^3_{\pi\gamma} (s)\over q^3_{\pi\gamma} (m_\omega)}
\Gamma_\omega B_{\omega\to\pi^0\gamma} +
{W_{\rho\pi} (s)\over W_{\rho\pi} (m_\omega)}
\Gamma_\omega B_{\omega\to 3\pi},\\
\Gamma_V(s)&={m_V^2\over s}{q^3_\pi (s)\over q^3_\pi (m_V)}\Gamma_V
 \mbox{~~~} (V =\rho,\rho^\prime).
\end{align}
Here, $q_{\pi\gamma}=(s-m^2_\pi)/2\sqrt{s}$, $W_{\rho\pi}(s)$ is the phase
space factor for the $\rho\pi\to\pi^+\pi^-\pi^0$ final state \cite{snd-3pi4},
$B_{V\to f}$ is the branching fraction of the vector meson decay to the final 
state $f$.  In the energy dependence of the $\rho$ and $\rho^\prime$ mesons widths 
only the $V\to\pi^+\pi^-$ decays are taken into account. Such approach is 
justified in the energy region $\sqrt{s}<1000$ MeV. 
The relative decay probabilities are calculated as follows
\begin{align}
B_{V\to f}&={\sigma(V\to f)\over\sigma(V)},\\
\sigma(V)&=\sum_{f} \sigma(V\to f),\\
\label{CS:Br}
\sigma(V\to f) &= {{12\pi B_{V\to e^+e^-}B_{V\to f} } \over {m_V^2}}.
\end{align}
The fit to the measured cross section $\sigma^0_{\pi\pi}$ is performed 
with the following 
free parameters $m_{\rho}$, $\Gamma_{\rho}$, $\sigma(\rho\to\pi^+\pi^-)$, 
$\sigma(\omega\to\pi^+\pi^-)$, $\sigma_(\rho^\prime\to\pi^+\pi^-)$ and 
$\phi_{\rho\omega}$. The values of $m_{\omega}$, $\Gamma_{\omega}$, 
$m_{\rho^\prime}$, $\Gamma_{\rho^\prime}$ are taken from \cite{pdg}. 
The relative phase $\phi_{\rho\rho^\prime} $ is fixed at $\pi$ according to 
ref.~\cite{snd2pi}. Only uncorrelated errors of the cross section are taken 
into account in the fit.
The results of the fit (figure~\ref{fig:pi2fit}) together with the results 
of the SND measurements at the VEPP-2M collider \cite{snd2pi-2} are presented in 
table~\ref{tab:p3}. The products 
\begin{equation} B_{V \to e^+e^-} \times B_{V\to\pi^+\pi^-} =
\frac{m_V^2\sigma(V\to\pi^+\pi^-)}{12\pi} \mbox{~~~} (V =\rho,\omega)
\end{equation}
are also presented in table~\ref{tab:p3}.

\begin{figure} [tbp]
\centering
\includegraphics [width = 0.8 \textwidth]{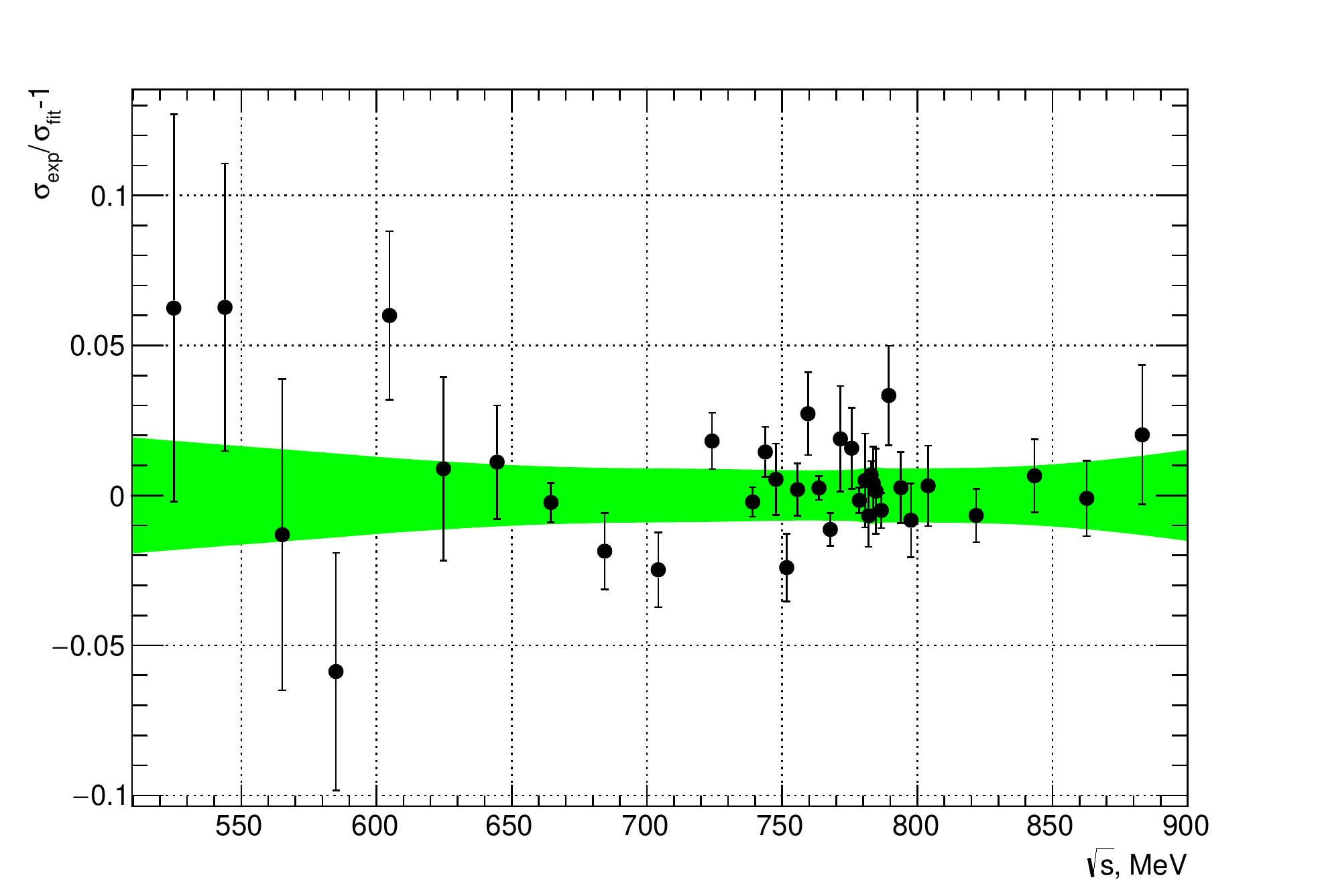}
\caption{The relative difference between the measured $e^+e^-\to\pi^+\pi^-$ cross 
section and the fit curve.}
\label{fitrel}
\end{figure}

The ratio between the measured cross section and the fit curve 
is shown in figure~\ref{fitrel}.
The systematic errors of $m_{\rho}$ and $\Gamma_{\rho}$ are related to the model 
uncertainty. It is estimated by comparison of the central values of these
parameters presented in table~\ref{tab:p3} with the results of the fit with a 
model based on the Gounaris-Sakurai parametrization \cite{kmd2-2,arbuzhad}.
If the $m_{\omega}$ and $\Gamma_{\omega}$ are free parameters
of the fit, their values are in agreement with those presented in 
PDG \cite{pdg}, and the value of $\phi_{\rho\omega}$ is shifted 
by $1^{\circ}$. This difference is taken as the systematic uncertainty of the 
$\phi_{\rho\omega}$.

\begin{table}
\begin{center}
\begin{tabular}{| c | c | c |} \hline
Parameter & This work & SND06 \\\hline
$m_\rho,$ MeV & 775.3 $\pm$ 0.5 $\pm$ 0.6 & 774.6 $\pm$ 0.4 $\pm$ 0.5 \\
$ \Gamma_\rho,$ MeV & 145.6 $\pm$ 0.6 $\pm$ 0.8 & 146.1 $\pm$ 0.8 $\pm$ 1.5 \\
$ \sigma(\rho \to \pi^+ \pi^-),$ nb & 1189.7 $\pm$ 4.5 $\pm$ 9.5 &
1193 $\pm$ 7 $\pm$ 16 \\
$\sigma(\omega\to\pi^+\pi^-)$, nb & 31.5 $\pm$ 1.2 $\pm$ 0.6 &
29.3 $\pm$ 1.4 $\pm$ 1.0 \\
$\phi_{\rho\omega}$, deg. & 110.7 $\pm$ 1.1 $\pm$ 1.0 &
113.7 $\pm$ 1.3 $\pm$ 2.0 \\
$\sigma(\rho^\prime\to\pi^+ \pi^-)$, nb & 2.4 $\pm$ 0.6 & 1.8 $\pm$ 0.2 \\
$\chi^2/ndf$ & 47/30 & -- \\
$B_{\rho \to e^+ e^-}\times B_{\rho\to\pi^+\pi^-}$ &
(4.889 $\pm$ 0.015 $\pm$ 0.039) $\times10^{-5}$ &
(4.876 $\pm$ 0.023 $\pm$ 0.064) $\times 10^{-5}$ \\
$B_{\omega\to e^+ e^-}\times B_{\omega\to\pi^+\pi^-}$ &
(1.318 $\pm$ 0.051 $\pm$ 0.021) $\times10^{-6}$ &
(1.225 $\pm$ 0.058 $\pm$ 0.041) $\times 10^{-6}$ \\\hline
\end{tabular}
\caption{Results of the fit obtained in this work together with results from 
ref.~\cite{snd2pi-2} (SND06). Both statistical and systematic errors are shown.}
\bigskip
\label{tab:p3}
\end{center}
\end{table}

\subsection{Contribution to the $ a_{\mu} $}

The contribution to the anomalous magnetic moment of the muon due to 
$\pi^+\pi^-(\gamma)$ intermediate state in the vacuum polarization is 
calculated via the dispersion integral
\begin{equation}
\label{amu}
a_\mu(\pi\pi, 525\mbox{MeV}\le\sqrt{s}\le 883\mbox{MeV}) =
\biggl ({\alpha m_\mu\over 3\pi}\biggr)^2\int^{S_{max}}_{S_{min}}
{R (s) K (s)\over s^2} ds,
\end{equation}
 where $K(s)$ is the known kernel \cite{g-2-1} and
\begin{align}
R(s)&={\sigma_{bare} \over \sigma(e^+ e^-\to\mu^+\mu^-)},\\
\sigma(e^+ e^-\to\mu^+\mu^-) &={4\pi\alpha^2\over 3s}.
\end{align}

Here $\sigma_{bare}$ (table~\ref{tab:pi2tab1}) is the bare cross section 
of the process $e^+e^-\to\pi^+\pi^-$ (the cross section without vacuum 
polarization contribution but taking into account the final state correction):
\begin{equation}
\label{spol}
\sigma_{bare}(s)=
\sigma^0_{\pi\pi}(s)\times|1-\Pi(s)|^2\times (1+\frac{\alpha}{\pi} a (s)),
\end{equation}
where $\Pi(s)$ is the polarization operator calculated according to the
ref.~\cite{arbuzqed} from the known $e^+e^-\to\mbox{hadrons}$ cross section
\cite{fedor}. The last factor $a(s)$ takes into account the final state 
radiation for the point-like pion \cite{shw}.

The integral (\ref{amu}) is evaluated by using the trapezoidal rule. 
As a result it is obtained
$$
 a_\mu (\pi\pi,525\mbox{MeV}\le\sqrt{s}\le 883\mbox{MeV}) = 
 (409.79 \pm 1.44 \pm 3.87)\times10^{-10}.
$$

The difference $1.8\times10^{-10}$ between this value and one, calculated 
with a fit curve, is taken as additional source of the systematics. 

\section{Discussion}

\begin{figure} [tbp]
\centering
\includegraphics [width = 0.8\textwidth]{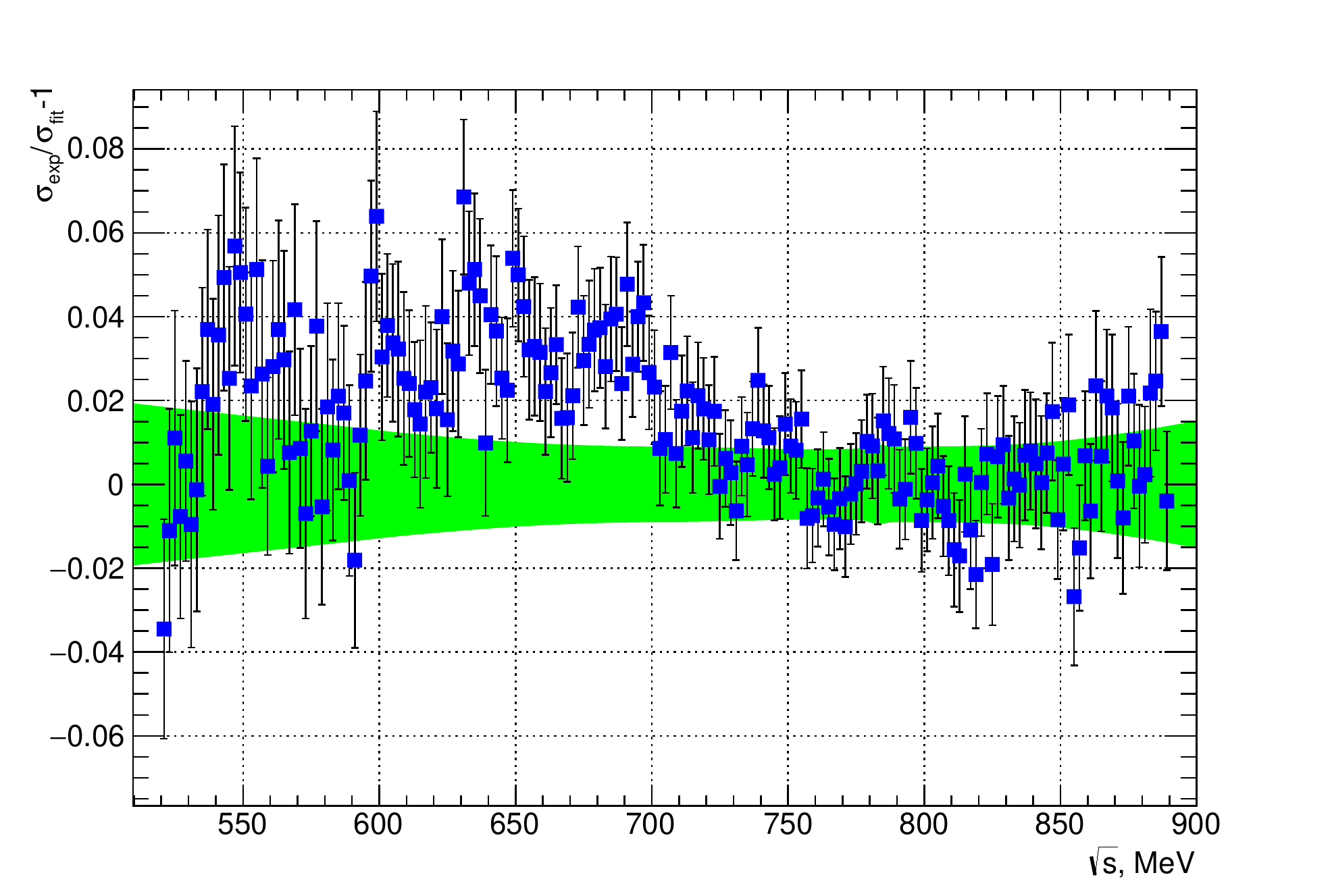}
\caption{The relative difference between the $e^+ e^-\to\pi^+\pi^-$ cross 
section measured by BaBar~\cite{babar} and the fit to the 
SND data. The error bars take into account both statistic and systematical 
errors of BaBar data. The shaded area corresponds to the quadratic sum of the 
systematic and statistical errors of the SND.}
\label{fig:snd_babar}
\end{figure}
\begin{figure} [tbp]
\centering
\includegraphics [width = 0.8\textwidth]{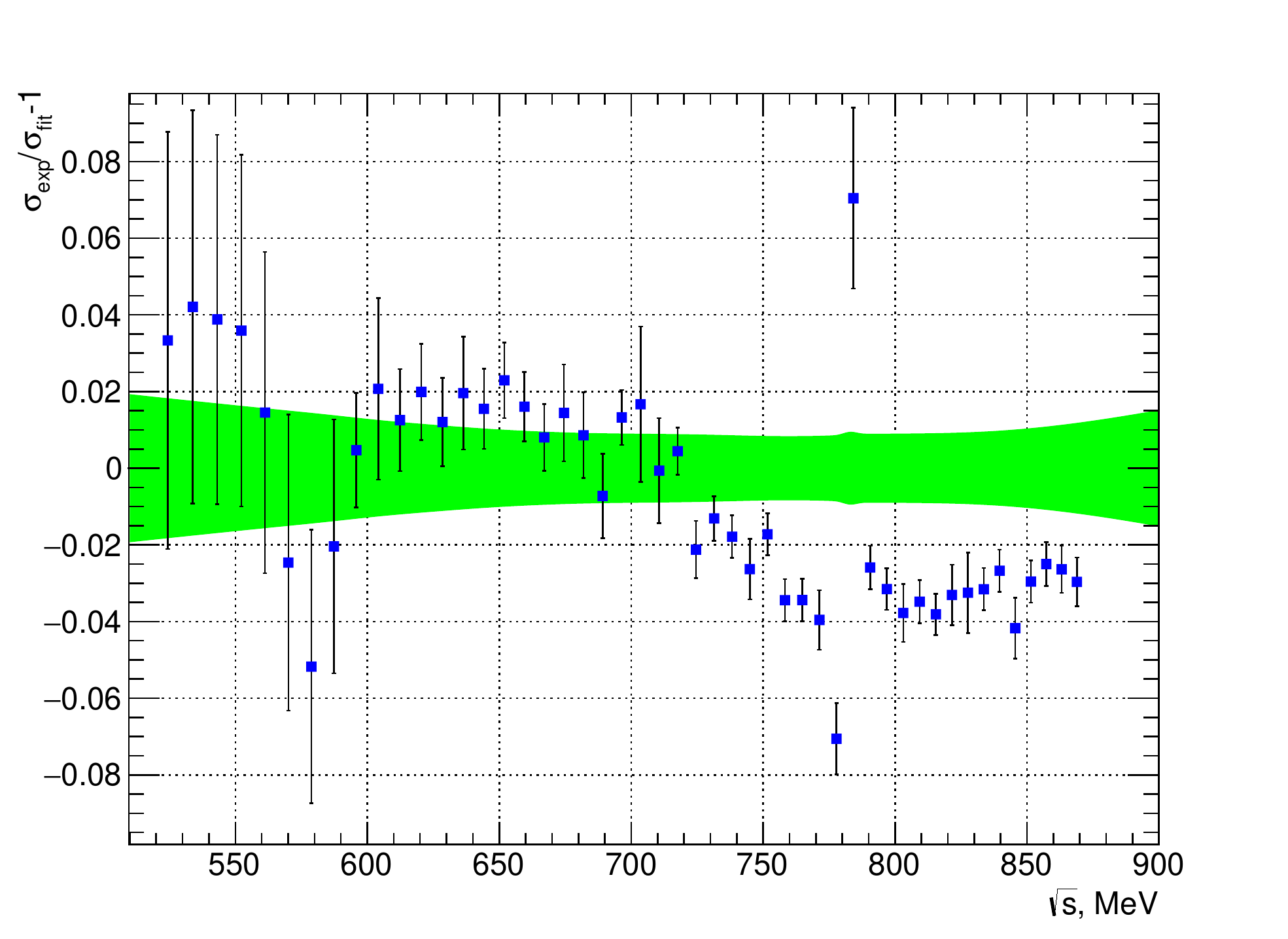}
\caption{The relative difference between the $e^+ e^-\to\pi^+\pi^-$ cross section 
measured by KLOE~\cite{kloe-2pi} and the fit to the SND data. The error 
bars take into account both statistic and systematical 
errors of KLOE data. The shaded area corresponds to the quadratic sum of the 
systematic and statistical errors of the SND. }
\label{fig:snd_kloe}
\end{figure}
\begin{figure} [tbp]
\centering
\includegraphics [width = 0.8\textwidth]{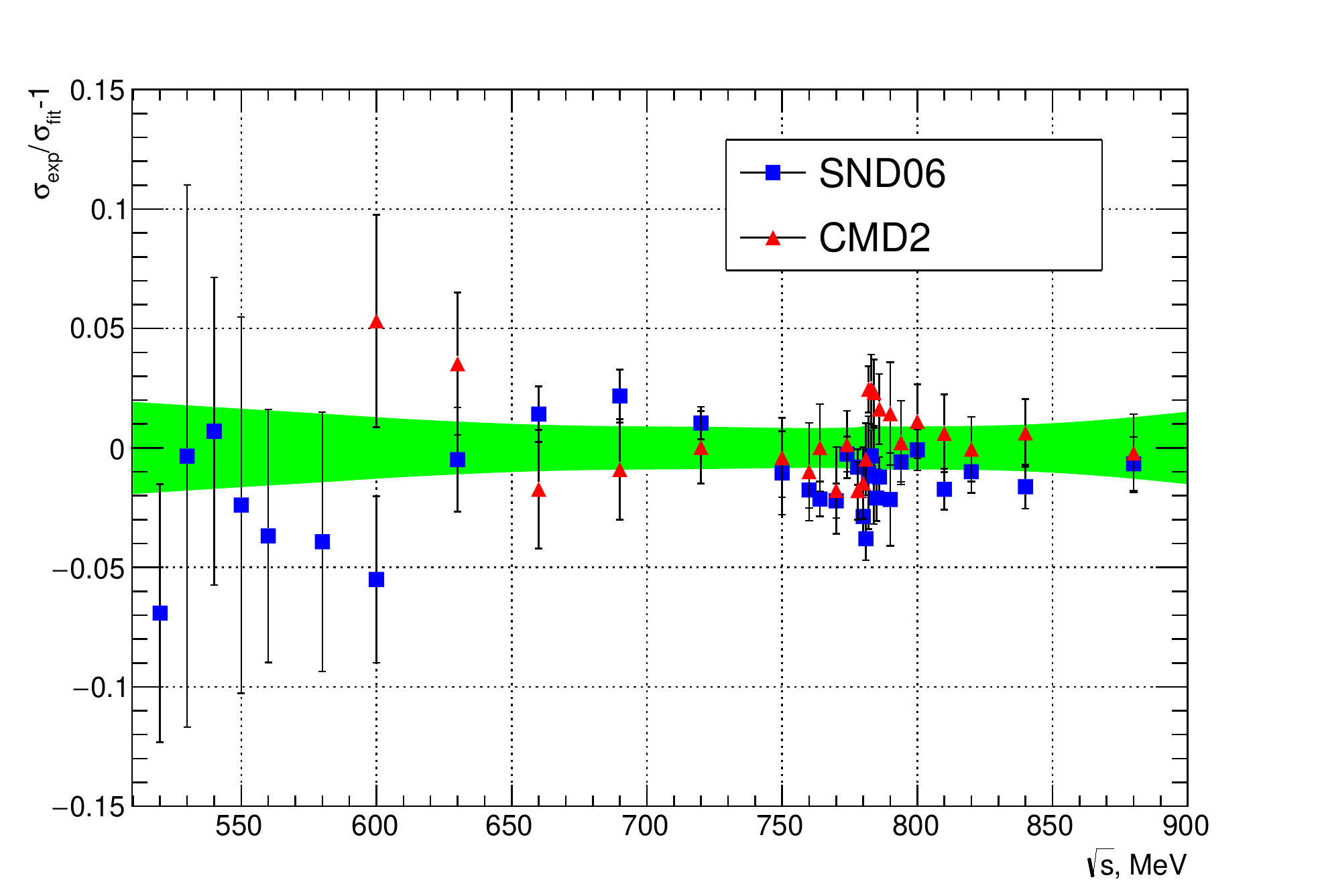}
\caption{The relative difference between the $e^+ e^-\to\pi^+\pi^-$ cross sections 
measured by SND~\cite{snd2pi-2} and CMD-2~\cite{kmd2-3} at VEPP-2M and the fit 
to the SND data at VEPP-2000. The error bars take into account both statistic
and systematical errors of VEPP-2M data. The shaded area corresponds to the 
quadratic sum of the systematic and statistical errors of the SND at
VEPP-2000.}
\label{fig:snds}
\end{figure}

The comparison of the $e^+e^-\to\pi^+\pi^-$ cross section obtained in this 
work with the results \cite{babar,kloe-2pi,snd2pi,kmd2-3} is
shown in figure~\ref{fig:snd_babar}, \ref{fig:snd_kloe}, \ref{fig:snds}.
The difference of 3\% between SND and BABAR data is observed in the energy 
region 0.62 $\ge\sqrt{s}\le$ 0.7 GeV, while outside it the SND and BABAR data 
are consistent (figure~\ref{fig:snd_babar}). The deviation between the KLOE and 
SND data is 1--3\% at $\sqrt{s}\ge$0.7 GeV. Below 0.7 GeV, the measurements are 
consistent (figure~\ref{fig:snd_kloe}). The results obtained in this work and 
in experiments at VEPP-2M are in agreement (figure~\ref{fig:snds}).


\begin{figure}
\centering
\begin{minipage} [t]{0.8\textwidth}
\includegraphics [width = 1.0\textwidth]{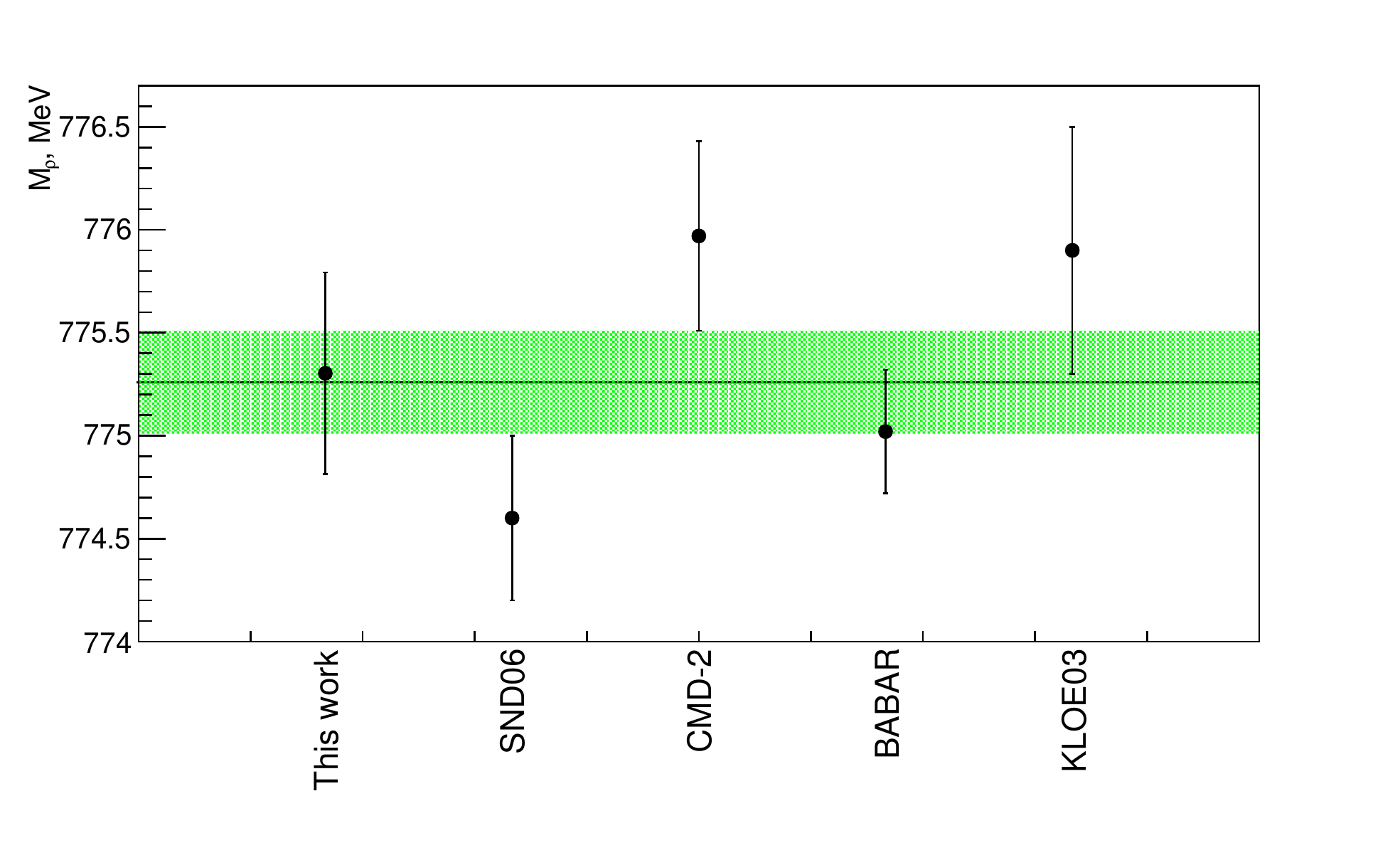}
\end{minipage}
\caption{The $\rho$ meson mass $m_{\rho}$ measured in this work and in refs.~\cite{snd2pi-2, kmd2-3, babar, kloe-3pi}. The shaded area shows the PDG value \cite{pdg}}
\label{M_rho}
\end{figure}

\begin{figure}
\centering
\begin{minipage} [t]{0.8\textwidth}
\includegraphics [width = 1.0\textwidth]{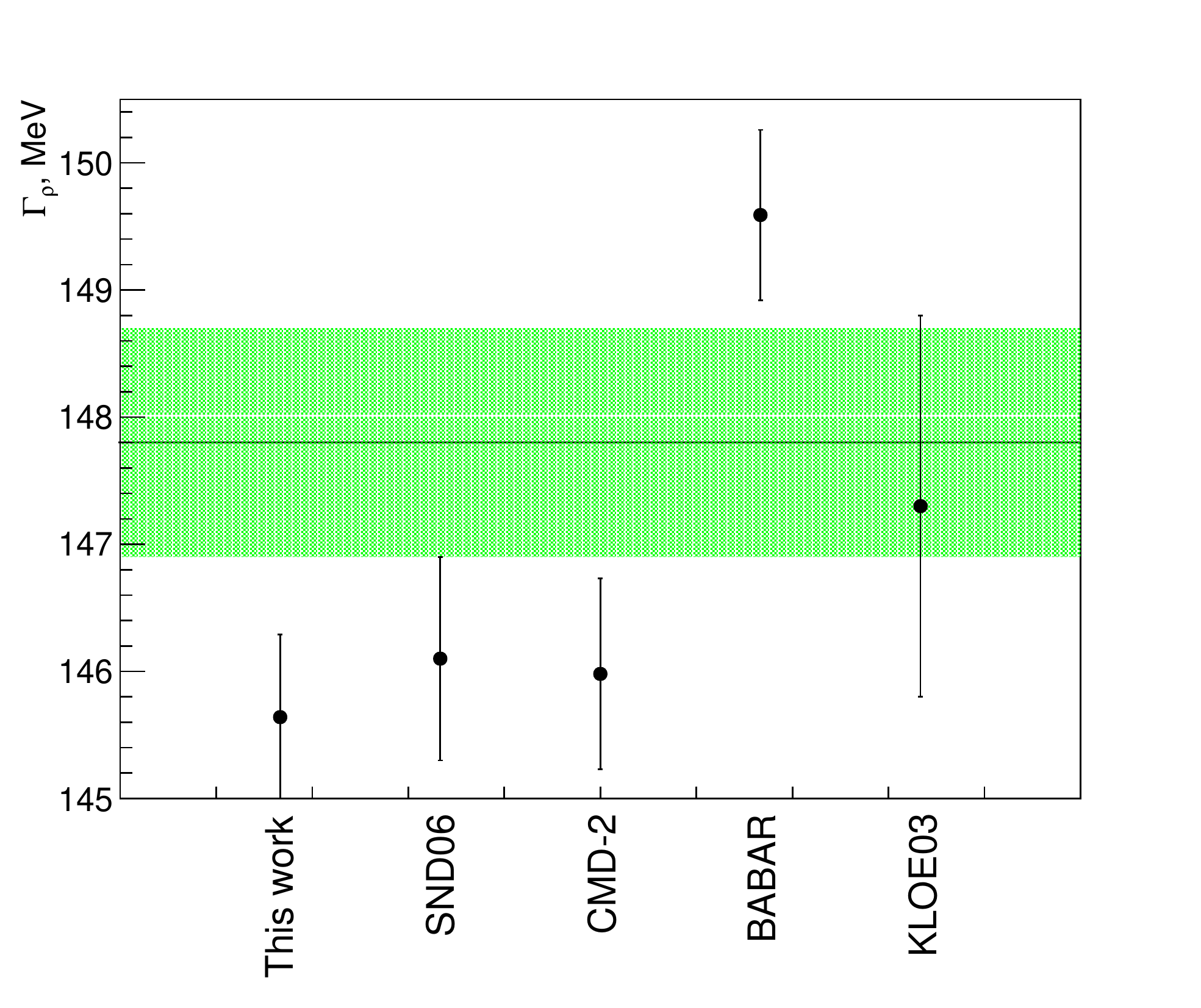}
\end{minipage}
\caption{The $\rho$ meson width $\Gamma_{\rho} $ measured in this work and in refs.~\cite{snd2pi-2, kmd2-3, babar, kloe-3pi}. The shaded area shows the PDG value \cite{pdg}}
\label{G_rho}
\end{figure}

The parameters of the $\rho$ and $\omega$ mesons obtained in this analysis 
are consistent with the results of \cite{snd2pi-2} (table~\ref{tab:p3} ). 
The $\rho$ meson mass $m_{\rho}$ is in agreement with the results of earlier 
experiments \cite{snd2pi-2, kmd2-3, babar, kloe-3pi} (figure~\ref{M_rho}).
Its width $\Gamma_{\rho}$ 
agrees with results of ref.~\cite{snd2pi-2, kmd2-3, kloe-3pi} and
contradict to the value reported by BaBar \cite{babar} (figure~\ref{G_rho}). To understand the source of the latter difference, we perform BABAR cross section fit in the energy region 0.525-0.883 GeV using our model (\ref{CS:S}). The obtained $\rho$ meson width is 147.38 MeV $\pm$ 0.47 MeV. We conclude that the discrepancy can be partially explained by difference between the fitting models.
 
The differences between 
$a_\mu(\pi\pi,525\mbox{MeV}\le\sqrt{s}\le 883\mbox{MeV})\times 10^{10}$ 
obtained in this work and those derived from \cite{snd2pi-2,babar} do 
not exceed one standard deviation, and there is a discrepancy between KLOE \cite{ kloe-2, kloe-3, kloe-4} and SND results (table~\ref{tab:amu}).

 
\begin{table}
\begin{center}
\begin{tabular}{|c|c|}\hline
Measurement & $a_\mu (\pi\pi)\times10^{10} $ \\\hline
This work & 409.79 $\pm $ 1.44 $\pm $ 3.87  \\
SND06 & 406.47 $\pm $ 1.74 $\pm $ 5.28 \\
BaBar & 413.58 $\pm $ 2.04 $\pm $ 2.29 \\
KLOE & 403.39 $\pm $ 0.72 $ \pm $ 2.50\\\hline
\end{tabular}
\caption{The contribution to the anomalous magnetic moment of the muon
$a_\mu(\pi\pi,525\mbox{MeV}\le\sqrt{s}\le 883\mbox{MeV})\times10^{10} $ 
derived from the SND and \cite{snd2pi-2, babar, kloe-2pi} data. The covariance matrix is 
used to calculate the statistical uncertainty for \cite{babar, kloe-2pi}.\bigskip}
 \label{tab:amu}
\end{center}
\end{table}

\section{Conclusion}

The cross section of the process $e^+e^-\to\pi^+\pi^-$ has been measured in the 
SND experiment at the VEPP-2000 collider in the energy region 
$525<\sqrt{s}<883 $ MeV. The systematic error of the measurement is 0.8\% 
at $\sqrt{s}>600$ MeV and 0.9--1.2 \% at $\sqrt{s}<600$ MeV. 
The measured cross section has been analyzed in the framework of the generalized 
vector meson dominance model. 
The following $\rho$ meson parameters have been obtained:
\begin{align*}
m_\rho &= 775.3\pm 0.5\pm 0.6 \mbox{~~MeV},\\
\Gamma_\rho &= 145.6\pm 0.6\pm 0.8 \mbox{~~MeV},\\ 
B_{\rho\to e^+e^-}\times B_{\rho\to\pi^+\pi^-}&=
(4.89\pm 0.02\pm 0.04)\times 10^{-5}.
\end{align*}
The parameters of the $G$-parity suppressed process 
$e^+e^-\to\omega\to\pi^+\pi^-$ has been measured: 
$$B_{\omega\to e^+ e^-}\times B_{\omega\to\pi^+\pi^-}=
(1.32\pm 0.06\pm 0.02)\times 10^{-6},$$
the relative phase between $\rho$ and $\omega$ mesons 
$$\phi_{\rho\omega} = (110.7\pm 1.5\pm1.0)^\circ.$$ 

The result of this work is in agreement with VEPP-2M measurements, but is 
in conflict with BaBar and KLOE measurements. The $\pi^+\pi^-(\gamma)$ 
contribution to the anomalous magnetic moment of the muon has been derived from 
the measured cross section: 
$a_\mu(\pi\pi,525\mbox{MeV}\le\sqrt{s}\le 883\mbox{MeV})=
(409.79 \pm 1.44 \pm 3.87)\times10^{-10}$.

\acknowledgments
The authors are grateful to B. Malaescu and A. Keshavarzi for useful discussions.
The work is supported in part by grants 
RFBR 18-02-00382-a, 18-02-00147-a, 20-02-00347-a, 20-02-00139-a,
20-02-00060-a.

\pagebreak


\begin{thebibliography}{99}
\bibitem{snd}
M.N.Achasov et al.,
\emph{Spherical neutral detector for VEPP-2M collider},
\emph{Nucl. Instr. Meth.} \textbf{A 449} (2000) 125 \\
\bibitem{snd-2}
M.N. Achasov et. al.,
\emph{Spherical Neutral Detector for experiments at VEPP-2000 $e^+e^-$
collider},
in proceedings of
\emph{International Workshop on ee collisions from Phi to Psi},
September 19 -- 22, 2011 Novosibirsk, Russia,
\emph{Nucl. Phys. Proc. Suppl.} \textbf{225-227} (2012) 66
\bibitem{vepp2k}
D.E. Berkaev et al.,
\emph{Electron-positron collider VEPP-2000. First experiments},
\emph{Zh. Eksp. Teor. Fiz.} \textbf{140} (2011) 247
\bibitem{g-2}
I. Logashenko et. al.,
\emph{The Measurement of the Anomalous Magnetic Moment of the Muon at
Fermilab},
\emph{J. Phys. Chem. Ref. Data} \textbf{44} (2015) 031211
\bibitem{cs2p}
I.B. Logashenko et al.,
\emph{Measurement of the hadronic cross sections at Novosibirsk},
in proceedings of
\emph{International Conference Dark Matter, Hadron Physics and Fusion
Physics},
September 24 -- 26, 2014 Messina, Italy,
\emph{EPJ Web Conf.} \textbf{96} (2015) 01022
\bibitem{kozev}
N. N. Achasov and A. A. Kozhevnikov \emph{Electromagnetic form factor of the pion in the field-theory-inspired approach}
\emph{Phys. Rev.} \textbf{D 85}, 019901 (2012)
\bibitem{augu}
J.E. Augustin et al., 
\emph{Study of Electron-Positron Annihilation into ${\ensuremath{\pi}}^{+}{\ensuremath{\pi}}^{\ensuremath{-}}$ at 775 MeV with the Orsay Storage Ring}, \emph{Phys. Rev. Lett.} {\bf 20}, 126, (1968)\\
\bibitem{augu-2}
J.E. Augustin et al., \emph{$\pi^+\pi^-$ production in $e^{+}e^{-}$ collisions and $\rho-\omega$ interference},
 in proceedings of 
 \emph{4th Rencontres de Moriond : Les Interactions Électromagnétiques}
 11-21 March 1969 Moriond, France, \emph{Nuovo Cim. Lett.} {\bf 2}, 214, (1969)\\
\bibitem{augu-3}
J.E. Augustin et al., \emph{Study of electron-positron annihilation into ${\ensuremath{\pi}}^{+}{\ensuremath{\pi}}^{\ensuremath{-}}$ on the $\varrho$o resonance}, \emph{Phys. Lett.} {\bf B 28}, 508, (1969)
\bibitem{ausl}
V.L. Auslender et al., \emph{Investigation of the $\varrho$-meson resonance with electron-positron colliding beams}, {Phys. Lett.} {\bf B 25}, 433, (1967) \\
\bibitem{ausl-2}
V.L. Auslander et al., \emph{Investigation of the rho-meson resonance with electron-positron colliding beams} \emph{Yad. Fiz.} {\bf 9}, 114, (1969) [\emph{Sov. J.Nucl. Phys.}
{\bf 9}, 69, 1969]
\bibitem{bena}
D. Benaksas et al., \emph{${\ensuremath{\pi}}^{+}{\ensuremath{\pi}}^{\ensuremath{-}}$ production by $e^{+}e^{-}$ annihilation in the $\varrho$ energy range with the Orsay storage ring}, \emph{Phys. Lett.} {\bf B 39}, 289, (1972)
\bibitem{quen}
A. Quenzer et al., \emph{Pion form factor from 480 MeV to 1100 MeV}, \emph{Phys. Lett.} {\bf B 76}, 512, (1978)
\bibitem{vas1}
I.B. Vasserman et al., \emph{Pion Form-Factor Measurement from $e+ e- --> pi+ pi-$ Near Threshold by electron-Positron Colliding Beams}, \emph{Yad. Fiz.} {\bf 28}, 968, (1978)
\bibitem{buki}
A.D. Bukin et al., \emph{Pion form factor measurement by $e^+ e^-\to\pi^+\pi^-$ in the energy range 2E from 0.78 up to 1.34 GeV}, \emph{Phys. Lett.} {\bf B 73}, 226, (1978)
\bibitem{vas2}
I.B. Vasserman et al., \emph{Measurement Of Pion Form-factor In $E+ E- ---> Pi+ Pi-$ Reaction Near Production Threshold}, \emph{Yad. Fiz.} {\bf 30}, 999, (1979) [\emph{Sov.J.Nucl.Phys.} \textbf{30} (1979) 519]
\bibitem{vas3}
I.B. Vasserman et al., \emph{Pion Form-factor Measurement in the Reaction $e^+ e^-\to\pi^+\pi^-$ for Energies Within the Range From 0.4-{GeV} to 0.46-{GeV}}, \emph{Yad. Fiz.} {\bf 33}, 709, (1981)
[\emph{Sov. J. Nucl. Phys.} {\bf 33}, 368, (1981)]
\bibitem{kur1}
L.M. Kurdadze et al., \emph{Measurement of the pion form factor at 640 $\leq \sqrt{s}\leq$ 1400 MeV}, \emph{JETP Lett.} {\bf 37}, 733, (1983)
[\emph{Pisma Zh. Eksp. Teor. Fiz.} {\bf 37}, 613, (1983)]
\bibitem{kur2}
L.M. Kurdadze et al., \emph{Study Of The Reaction $E+ E- ---> Pi+ Pi-$ In The Energy Range From 640-mev - 1400-mev}, \emph{Yad. Fiz.} {\bf 40}, 451, (1984)
[\emph{Sov. J. Nucl. Phys.} {\bf 40}, 286, (1984)]
\bibitem{spec}
S.R. Amendolia et al., \emph{ Measurement of the pion form factor in the time-like region for $q^2$ values between 0.1 $(GeV/c)^2$ and 0.18 $(GeV/c)^2$}, \emph{Phys. Lett.} {\bf B 138}, 454, (1984)
\bibitem{olya}
L.M. Barkov, et al., \emph{Electromagnetic pion form factor in the timelike region}, \emph{Nucl. Phys. B} {\bf 256}, 365, (1985)
\bibitem{kmd2}
R.R. Akhmetshin et al., \emph{Measurement of $e^+e^-\to\pi^+\pi^-$ cross-section with CMD-2 around $\rho$-meson}, \emph{Phys. Lett.} {\bf B 527}, 161, (2002) [arXiv:hep-ex/0112031] \\
\bibitem{kmd2-1}
R.R. Akhmetshin et al., \emph{Update: A reanalysis of hadronic cross section measurements at CMD-2}, \emph{Phys. Lett.} {\bf B 578}, 285, (2004) [arXiv:hep-ex/0308008]
\bibitem{kloe}
A. Aloisio et al., \emph{Measurement of $\sigma(e^+ e^-\to\pi^+\pi^-\gamma)$ and extraction of $\sigma(e^+ e^-\to\pi^+\pi^-)$ below 1 GeV with the KLOE detector}, \emph{Phys. Lett.} {\bf B 606}, 12, (2005) [arXiv:hep-ex/0407048]
\bibitem{snd2pi}
M.N. Achasov et. al., \emph{Study of the process $e^+e^-\to\pi^+\pi^-$ in the energy region 400$<\sqrt{s}<$1000 MeV}, \emph{J.Exp.Theor.Phys.} \textbf{101} (2005) no.6, 1053-1070, [\emph{Zh.Eksp.Teor.Fiz.} \textbf{128} (2005) no.6, 1201-1219] [arXiv:hep-ex/0506076]\\
\bibitem{snd2pi-2}
M.N. Achasov et.al., \emph{Update of the $e^+e^-\to\pi^+\pi^-$ cross section measured by the spherical neutral detector in the energy region 400 $<\sqrt{s}<$ 1000 MeV}, \emph{J.Exp.Theor.Phys.} \textbf{103} (2006) 380-384 [\emph{Zh.Eksp.Teor.Fiz.} \textbf{130} (2006) 437-441][arXiv:hep-ex/0605013]
\bibitem{kmd2-2}
V.M. Aulchenko et. al., \emph{Measurement of the $e^+ e^-\to\pi^+\pi^-$ cross section with the CMD-2 detector in the 370–520-MeV energy range}, \emph{JETP Lett.} \textbf{84} (2006) 413-417, 
[\emph{Pisma Zh.Eksp.Teor.Fiz.} \textbf{84} (2006) 491-495][arXiv:hep-ex/0610016]
\bibitem{kmd2-3}
R.R. Akhmetshin et al, \emph{High-statistics measurement of the pion form factor in the $\rho$-meson energy range with the CMD-2 detector}, \emph{Phys.Lett.} \textbf{B 648} (2007) 28-38 [arXiv:hep-ex/0610021]
\bibitem{kloe-2}
A. Ambrosino et. al., \emph{Measurement of $\sigma(e^+ e^-\to\pi^+\pi^-\gamma(\gamma))$ and the dipion contribution to the muon anomaly with the KLOE detector}, \emph{Phys. Lett.} \textbf{B 670} (2009) 285-291 [arXiv:0809.3950]
\bibitem{kloe-3}
F. Ambrosino et. al., \emph{Measurement of $\sigma(e^+ e^-\to\pi^+\pi^-)$ from threshold to 0.85 $GeV^2$ using initial state radiation with the KLOE detector}, \emph{Phys. Lett.} \textbf{B 700} (2011) 102-110 [arXiv:1006.5313]
\bibitem{kloe-4}
A. Anastasi et. al., \emph{Measurement of the running of the fine structure constant below 1 GeV with the KLOE Detector}, \emph{JHEP} \textbf{1803}(2018) 173 [arXiv:1609.06631]
\bibitem{babar}
B. Aubert et. al., \emph{Precise Measurement of the $e^+ e^-\to\pi^+\pi^-\gamma$ Cross Section with the Initial State Radiation Method at BABAR}, \emph{Phys. Rev. Lett.} \textbf{103} (2009) 231801 \\
\bibitem{babar-2}
J.P. Lees et. al., \emph{Precise measurement of the $e^+ e^-\to\pi^+\pi^-\gamma$ cross section with the initial-state radiation method at BABAR}, \emph{Phys. Rev.} \textbf{D 86} (2012) 032013 [arXiv:1205.2228]
\bibitem{bes3}
M. Ablikim et. al., \emph{Measurement of the $e^+ e^-\to\pi^+\pi^-$ cross section between 600 and 900 MeV using initial state radiation}, \emph{Phys.Lett.} \textbf{B 753} (2016) 629-638 [arXiv:1507.08188]
\bibitem{ashif}
A.Yu. Barnyakov et. al., \emph{Testing aerogel Cherenkov counters with
n = 1.05 using electrons and muons},
\emph{Prib.Tekh.Eksp.} \textbf{1} (2015) 37
[\emph{Instrum.Exp.Tech.} \textbf{58} (2015) 30]
\bibitem{ems}
E.V. Abakumova et al.,
\emph{A system of beam energy measurement based on the
Compton backscattered laser photons for the VEPP-2000 electron-positron
collider},
\emph{Nucl. Instr. Meth.} \textbf{A 744} (2014) 35-40 [arXiv:1310.7764]\\
\bibitem{ems-2}
E.V. Abakumova et al., \emph{Backscattering of Laser Radiation on Ultrarelativistic Electrons in a Transverse Magnetic Field: Evidence of MeV-Scale Photon Interference}, \emph{Phys.Rev.Lett.} \textbf{110} (2013) no.14, 140402 [arXiv:1211.0103] 
\bibitem{geant4}
S. Agostinelli et. al., \emph{Geant4--a simulation toolkit}, \emph{Nucl. Instr. Meth.} \textbf{A 506}, 250 (2003) \\
\bibitem{geant4-2}
J. Allison et. al., \emph{Geant4 developments and applications}, \emph{IEEE Trans. on Nucl. Science} \textbf{53}, 270 (2006)
\bibitem{mcgpj}
G.V. Fedotovich, A.I. Sibidanov, \emph{Monte Carlo generator with radiative corrections for the $e^+e^-\to e^+e^-,\mu^+\mu^-$ and $\pi^+\pi^-$ processes at low energies}, \emph{Nuc. Phys. B Proc. Suppl.} \textbf{131} (2004) 9-18
\bibitem{arbuzqed}
A.B. Arbuzov et al., \emph{Large angle QED processes at $e^+e^-$ colliders at energies below 3 GeV}, \emph{JHEP} \textbf{9710}, 001 (1997) [arXiv:hep-ph/9702262]
\bibitem{arbuzhad}
A.B. Arbuzov et al., \emph{Radiative corrections for pion and kaon production at $e^+e^-$ colliders of energies below 2 GeV}, \emph{JHEP} \textbf{9710}, 006 (1997) [arXiv:hep-ph/9703456]
\bibitem{epi}
M.N. Achasov, K.I. Beloborodov and A.S. Kupich, \emph{Separation of $e^+ e^-\to e^+e^-$ and $e^+ e^-\to\pi^+\pi^-$ events using SND detector calorimeter}, \emph{JINST} \textbf{12T} 01002 [arXiv:1611.07729]
\bibitem{KOOP}
R. R. Akhmetshin et al., \emph{Search for the process $e^+ e^-\to \eta^{\prime}(958)$ with the CMD-3 detector}, \emph{Phys. Lett.B} \textbf{740}, 273 (2015) [arXiv:1409.1664]
\bibitem{snd-3pi4}
M.N. Achasov, K.I. Beloborodov, A.V. Berdyugin et. al., 
\emph{Study of the process $e^+e^-\to\pi^+\pi^-\pi^0$ in the energy region 
$\sqrt{s}$ below 0.98 GeV},
\emph{Phys. Rev.} (2003) Vol. \textbf{68D}, 052006 [arXiv:hep-ex/0305049]
\bibitem{pdg}
M. Tanabashi et al. (Particle Data Group)
\emph{Review of Particle Physics,}
\emph{Phys. Rev.} {\bf D 98}, 030001 (2018). 
 \bibitem{g-2-1}
K.Hagiwara, R.Liao, A.D.Martin, D.Nomura and T.Teubner.
\emph{$(g-2)_\mu$ and $\alpha(M_Z^2)$ re-evaluated using new precise data.}
\emph{J. Phys. G} {\bf 38}, 085003 (2011) [arXiv:1105.3149]
\bibitem{fedor}
F. Ignatov https://cmd.inp.nsk.su/$\sim$ignatov/vpl/
\bibitem{shw}
 \emph{J.Schwinger, Particles, Sources and Fields, vol.II,}
 Addison-Wesley Publishing Company Advanced Book Program Reading, Massachusetts, 1973
\bibitem{kloe-2pi}
A. Anastasi et al., \emph{Combination of KLOE $\sigma(e^+ e^-\to\pi^+\pi^-\gamma(\gamma))$ measurements and determination of $a_{\mu}^{\pi^+\pi^-}$ in the energy range 0.10 $< s <$ 0.95 $GeV^2$}, \emph{JHEP} \textbf{1803} (2018) 173 [arXiv:1711.03085]
\bibitem{kloe-3pi}
A. Aloisio et al., \emph{Study of the decay $\phi\to\pi^+\pi^-\pi^0$ with the KLOE detector}, \emph{Phys.Lett.} \textbf{B 561} (2003) 55-60 [arXiv:hep-ex/0303016]
\end{thebibliography}
\end{document}